\documentclass[twocolumn,secnumarabic,amssymb, nobibnotes, aps, prb, floatfix, showpacs]{revtex4}
\usepackage{graphicx}

\begin{document}

\title{Thermal fluctuations in moderately damped Josephson junctions: Multiple escape and retrapping, switching- and return-current distributions and hysteresis}

\author{J.~C.~Fenton}
\email[]{j.fenton@ucl.ac.uk}
\author{P.~A.~Warburton}
\affiliation{London Centre for Nanotechnology, 17--19 Gordon Street,
London WC1H 0AH, UK} \affiliation{UCL, Department of Electronic \&
Electrical Engineering, Torrington Place, London WC1E 7JE, UK.}

\date{\today}

\begin{abstract}
A crossover at a temperature $T^*$ in the temperature dependence of
the width $\sigma$ of the distribution of switching currents of
moderately damped Josephson junctions has been reported in a number
of recent publications, with positive
$\textrm{d}\sigma{}/\textrm{d}T$ and $IV$ characteristics associated
with underdamped behaviour for lower temperatures $T<T^*$, and
negative $\textrm{d}\sigma{}/\textrm{d}T$ and $IV$ characteristics
resembling overdamped behaviour for higher temperatures $T>T^*$. We
have investigated in detail the behaviour of Josephson junctions
around the temperature $T^*$ by using Monte Carlo simulations
including retrapping from the running state into the supercurrent
state as given by the model of Ben-Jacob \textit{et al.}
We develop discussion of the important role of multiple escape and
retrapping events in the moderate-damping regime, in particular
considering the behaviour in the region close to $T^*$. We show that
the behaviour is more fully understood by considering \textit{two}
crossover temperatures, and that the shape of the distribution and
$\sigma(T)$ around $T^*$, as well as at lower $T<T^*$, are largely
determined by the shape of the conventional thermally activated
switching distribution.  We show that the characteristic
temperatures $T^*$ are not unique for a particular Josephson
junction, but have some dependence on the ramp rate of the applied
bias current.
We also consider hysteresis in moderately damped Josephson junctions
and discuss the less commonly measured distribution of return
currents for a decreasing current ramp.
 We find that some hysteresis should be expected to persist above
$T^*$ and we highlight the importance, even well below $T^*$, of
accounting properly for thermal fluctuations when determining the
damping parameter $Q$.

[Accepted for publication in PRB; \copyright American Physical
Society 2008]
\end{abstract}

 \maketitle

\section{Introduction}
The Josephson junction system has been extensively studied both
theoretically and experimentally. Theoretically it has been
considered a model system for studying escape from a metastable
potential well. Experimentally, Josephson junctions have found
numerous applications and are presently being used in several
quantum bit implementations. In such experiments, an understanding
of the influence of thermal fluctuations is crucial in developing
applications.  Josephson junctions can be characterised by a damping
parameter $Q$. The majority of the large body of previous work in
the literature has concentrated on junctions in either the
underdamped ($Q\gg{1}$) or overdamped ($Q\sim{1})$ limits. In this
paper, we focus on the intermediate ``moderately damped''
($Q\approx{5}$) limit, where thermal fluctuations lead to
interesting physical effects.

For strongly underdamped Josephson junctions under the influence of
thermal fluctuations, the $IV$ characteristics are hysteretic and
the dynamics of switching from the zero-voltage supercurrent state
to the finite-voltage resistive phase-slip state are well described
by the analysis of Fulton and Dunkleberger\cite{fulton}, with a
distribution in switching currents as a result of thermal
fluctuations.  In contrast, overdamped junctions show non-hysteretic
behaviour, with a finite voltage on the supercurrent branch of the
$IV$ characteristic, associated with thermally activated phase
diffusion, and thermal fluctuations leading to very much smaller
variations in the switching behaviour. Phase diffusion in junctions
with hysteretic $IV$ characteristics has been discussed by Kautz and
Martinis\cite{kautz} and is associated with frequency-dependent
damping, such that junctions are underdamped at low frequencies, but
in the overdamped limit at high frequency.

The temperature dependence of the switching current and the width of
its distribution are experimental parameters of much recent
interest.
Experimental evidence of a crossover in the temperature dependence
of the switching current was reported first by Franz \textit{et
al.}\cite{franz} in experiments on small ``intrinsic'' Josephson
junctions (IJJs). They obtained $IV$ curves characteristic of
underdamped junctions below a crossover temperature and $IV$ curves
characteristic of overdamping above that temperature.
More recent experimental papers have reported a crossover in the
temperature dependence of the width $\sigma$ of the switching
current\cite{kivioja,mannik,kras1} at a temperature $T^*$, with
positive d$\sigma/\textrm{d}T$ below $T^*$ and negative
d$\sigma/\textrm{d}T$ above $T^*$. This was associated with a regime
of moderate damping. The negative d$\sigma/\textrm{d}T$ region was
associated with retrapping of the phase following escape. The
low-temperature behaviour fits the expectations for underdamped
junctions, and the high-temperature behaviour resembles previous
observations for overdamped junctions with phase diffusion. One
might simply explain the crossover from underdamped to overdamped
behaviour by a temperature-dependent damping $Q$ and this was indeed
the suggestion of Franz \textit{et al.} However, it was demonstrated
by Krasnov \textit{et al.}\cite{kras1} that such a crossover should
also be expected even for temperature-independent $Q$ if the
junctions are in the moderately damped regime ($Q\sim{}5$). Krasnov
\textit{et al.} derived an approximate quantitative formula with
$T^*=T^*(Q)$, implying that $T^*$ is a measure of the damping.

Several theoretical treatments of the retrapping process have been
presented\cite{benjacob,chen,cristiano} and the analysis of
retrapping was conducted in various ways in the experimental reports
of a crossover in $\sigma(T)$. In the analysis of
Ref.~\onlinecite{kivioja}, retrapping was assumed to be determined
purely by energetic considerations: retrapping is certain to occur
where it is energetically expected, below a current
$I_\textrm{m}\approx{}4I_\textrm{c}/\pi{}Q$, where $I_\textrm{c}$ is
the (fluctuation-free) critical current of the Josephson junction.
For $I>I_\textrm{m}$ there is an energy cost $\Delta{U}_\textrm{R}$
to retrapping --- in Ref.~\onlinecite{kivioja}, retrapping was
neglected for $I>I_\textrm{m}$. Krasnov \textit{et al.}\cite{kras1}
treated retrapping above $I_\textrm{m}$ as a thermally activated
process, with an energy barrier $\Delta{U}_\textrm{R}$, using the
model of Ben-Jacob \textit{et al.}\cite{benjacob} M\a"annik
\textit{et al.}\cite{mannik} used Monte Carlo simulations with an
RCSJ model including frequency-dependent damping to determine the
probability of thermally induced retrapping following escape.

In this article, in order to conduct a semi-analytic analysis of the
multiple escape and retrapping processes, we have adopted the model
of Ben-Jacob \textit{et al.}  We also include the effects of
frequency-dependent damping (see Section \ref{freqdepdt}). We
develop discussion of the important role of multiple escape and
retrapping events in the moderate-damping regime and present results
of Monte Carlo simulations showing the variation with experimental
parameters of the mean and width of the switching-current
distribution.\cite{fenton2} We consider the crossover between the
lower-temperature conventional underdamped regime and the
higher-temperature overdamped regime. Although previous studies have
considered a single crossover temperature, we show that, in detail,
the change occurs in two stages, with a lower-temperature transition
from underdamped behaviour to behaviour in the crossover regime, and
a higher temperature transition from the crossover regime to the
higher-temperature overdamped regime.
We demonstrate a significant change in the shape of the switching
current distribution around the crossover and study this
quantitatively through the skewness parameter.

The process of return from the resistive state in a hysteretic
junction is a much less well-studied phenomenon than that of escape.
Here we also consider the process of return from the resistive state
to the supercurrent state as the current is ramped down, and the
resulting variation in hysteresis around $T^*$. We compare our
findings with previous reports in the literature.

In our Monte Carlo simulations, for a current $I$, the probability
in a short time interval $\delta{t}$ of a transition between the
metastable and running states is given by
$\Gamma_\textrm{E}(I)\delta{t}$ for escape from the metastable state
(with $\Gamma_\textrm{E}$ given below by Eqn.~\ref{GE}) or
$\Gamma_\textrm{R}(I)\delta{t}$ for retrapping from the running
state (with $\Gamma_\textrm{R}$ given below by Eqn.~\ref{GR}).  A
bias current is ramped up (or down) at a constant rate in order to
generate distributions of switching (or return) currents for
junctions with a number of different parameters. We neglect the
temperature dependence of the critical current $I_\textrm{c}$ and
$Q$ in order to emphasize effects due to thermal fluctuations in the
junctions. As a bias current is ramped up, a switch is counted when
the junction spends more than half the time in the running state
over some time period $\tau$.\footnote{This period would be set by
the details of an experiment
--- see Section \ref{freqdepdt}.} Throughout this article, we use
the term ``escape'' to describe any (possibly short-lived) escape
from the instantaneous zero-voltage state, and reserve the term
``switch'' to describe an experimentally measured switch to the
running state. Similarly, when describing the behaviour as an
applied current is ramped \textit{down} from the critical current,
we reserve the term ``retrapping'' to describe a (possibly
short-lived) change from the voltage state to the zero-voltage
state, and use the term ``return'' to describe an experimentally
measured change from the voltage state to the zero-voltage state.

\subsection{The RCSJ model - the underdamped regime}\label{ud}

For a resistively shunted Josephson junction in the absence of
fluctuations, escape from the supercurrent state to a state of
finite voltage characterized by the junction resistance occurs when
the current bias applied to the junction reaches the junction
critical current $I_\textrm{c}$.
At finite temperatures, thermal fluctuations lead to switching at
currents below $I_\textrm{c}$, and there arises experimentally a
distribution in possible values of the switching current.  A common
experimental configuration is to ramp the current up from zero at a
constant rate $\textrm{d}I/\textrm{d}t$.  In that case, the
probability of a switch in the current range $I$ to $I+\textrm{d}I$
is $p(I)\textrm{d}I$, with\cite{fulton}
\begin{equation} \label{pI}
p(I)=\frac{\Gamma_\textrm{E}}{\textrm{d}I/\textrm{d}t}\bigg[1-\int_0^I
p(I')\textrm{d}I'\bigg],
\end{equation}
where $\Gamma_\textrm{E}$ is the rate, at current $I$, of escape
from the supercurrent state.

A mechanical analog for the resistively and capacitively shunted
Josephson junction (RCSJ) is that of a particle in a washboard
potential; it is often used in discussing the dynamics of such
junctions. \cite{tinkham} The height of the corrugations in the
untilted washboard is set by the Josephson energy. The current bias
corresponds to a tilt of the washboard, and position of the particle
along the washboard corresponds to the phase difference across the
junction, so that the speed of the particle as it moves in the
washboard potential corresponds to the voltage across the junction.
As it moves along the washboard, the particle is subject to a
viscous damping force which is inversely proportional to the
resistance shunting the junction.  The strength of the damping can
be characterized by a quality factor parameter\footnote{The McCumber
parameter $\beta_c\equiv{Q^2}$ is also sometimes used to
characterize the damping.} $Q=\omega_\textrm{P}RC$, where $R$ and
$C$ are the resistance and capacitance shunting the junction and
$\omega_\textrm{P}=\sqrt{2eI_\textrm{c}/\hbar{}C}$ is the angular
frequency of small oscillations at the bottom of the potential well
at zero bias. Hysteretic $IV$ characteristics are obtained for $Q\gg
1$ and phase diffusion obtained for $Q\sim 1$.

\subsection{Characteristic rates}
\begin{figure}[!hbp]
\includegraphics[width=8.6cm]{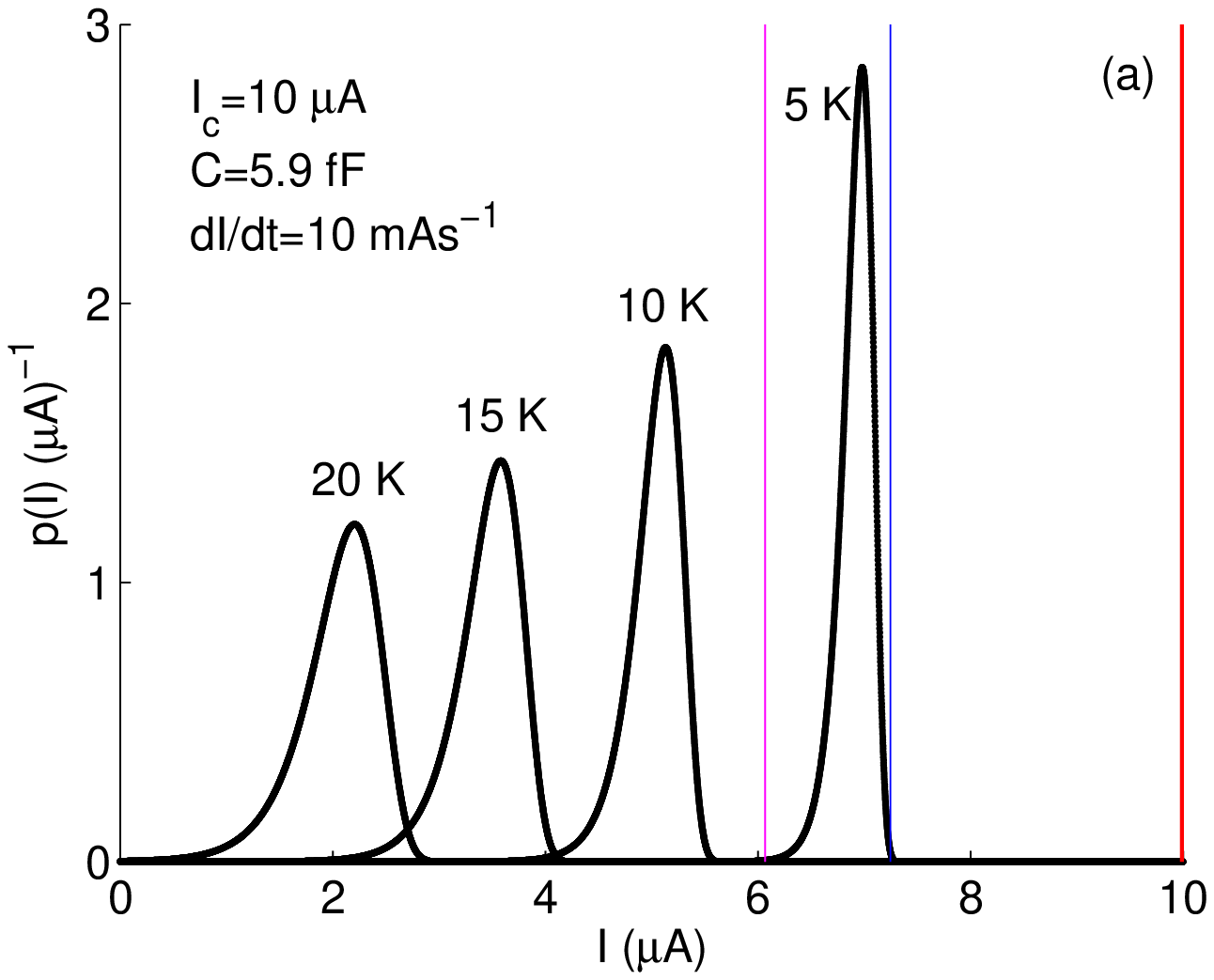}
\includegraphics[width=8.6cm]{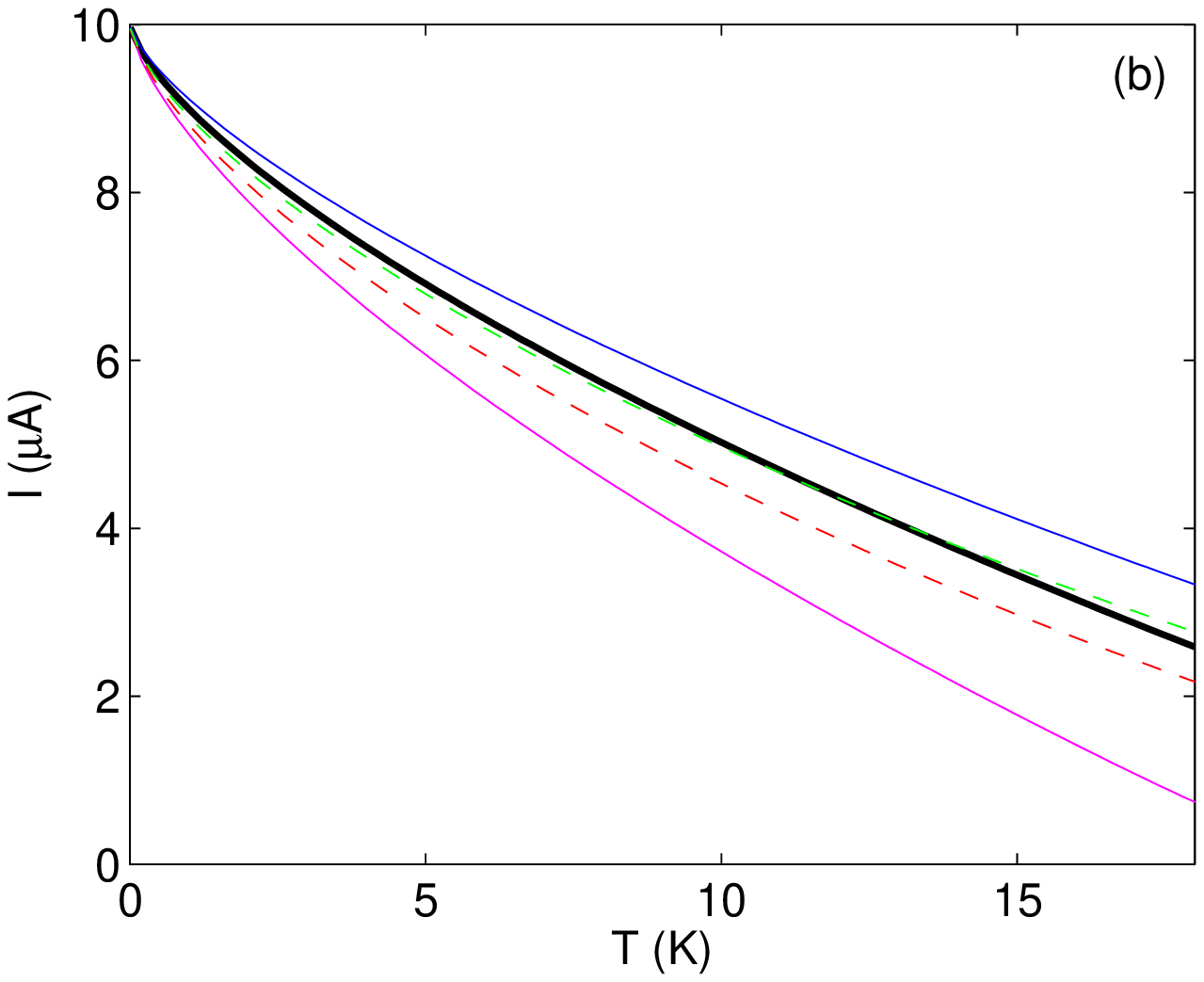}
\caption{(Color online) (a)\label{fig:pIt} Calculated underdamped
thermally activated $p(I)$ distributions at a number of
temperatures.  The parameters indicated are used for subsequent
figures unless otherwise stated; these parameters might be typical
for an IJJ.\cite{fenton} Vertical lines (pink and blue online) show
the boundaries of the distribution at 5 K, within which 99.99\% of
switching events occur. The thick vertical line (red online) shows
the critical current. (b)\label{fig:IswthT} Variation with
temperature of the mean switching current (thick black line) and the
top (blue online) and bottom (pink online) of the switching current
distribution. The lower and upper broken lines (red and green
online) show respectively $I(\Gamma_\textrm{E}=\Gamma_\textrm{I})$
and $I(\Gamma_\textrm{E}=10\Gamma_\textrm{I}$).}
\end{figure}
The rate of thermally activated escape from a minimum in the
washboard potential is given by \cite{kramers}
\begin{equation} \label{GE}
\Gamma_\textrm{E}=a_\textrm{t}\frac{\omega_\textrm{a}}{2\pi}\exp{\bigg(-\frac{\Delta{U}_\textrm{E}}{kT}\bigg)},
\end{equation}
where $\Delta{U}_\textrm{E}$ is the height of the energy barrier
from a washboard potential minimum to the adjacent maximum,
$a_\textrm{t}$ is a damping-dependent pre-factor and the quantities
$a_\textrm{t}, \omega_\textrm{a}$ and $\Delta{U}_\textrm{E}$ are all
current dependent, with
$\omega_\textrm{a}=\omega_\textrm{P}(1-(I/I_\textrm{c})^2)^{1/4}$
and, close to $I_\textrm{c}$,
$\Delta{U}_\textrm{E}\approx{}\frac{4\sqrt{2}}{3}E_\textrm{J}(1-I/I_\textrm{c})^{3/2}$
where the Josephson energy
$E_\textrm{J}={\hbar{I_\textrm{c}}}/{2e}$. Combining Eqns.~\ref{pI}
and \ref{GE} gives a characteristic asymmetric distribution of
switching currents for such junctions,
 as shown in Fig.~\ref{fig:pIt}a.

In the underdamped regime, the mean switching current decreases as
the temperature increases (Fig.~\ref{fig:IswthT}b) because larger
thermal fluctuations enable escape from the washboard minimum at a
lower current. The width of the switching current distribution may
be shown to depend on temperature as $\sigma \sim T^{2/3}$.

Thermal fluctuations can also cause retrapping of a particle which
has escaped from a potential well.  Ben-Jacob \textit{et al.}
\cite{benjacob} obtained an analytic formula for the rate
$\Gamma_\textrm{R}$ of this retrapping in the limit $Q\gg 1$. The
retrapping rate is strongly dependent on the damping through $Q$,
and is given by
\begin{equation} \label{GR}
\Gamma_\textrm{R}=\frac{I-I_\textrm{r}}{I_\textrm{c}}\omega_\textrm{P}\sqrt{\frac{E_\textrm{J}}{2\pi{}kT}}\exp{\bigg[-\frac{E_\textrm{J}Q^2(I-I_\textrm{r})^2}{2kTI_\textrm{c}^2}\bigg]},
\end{equation}
where $I_\textrm{r}=I_\textrm{r}(Q)\approx{4I_\textrm{c}}/{\pi{}Q}$.
Rewriting this in the form
$\Gamma_\textrm{R}\sim\exp{(-\Delta{U}_\textrm{R}/kT)}$ defines an
energy barrier $\Delta{U}_\textrm{R}$ for retrapping.\cite{kras1}
Eqn.~\ref{GR} has been applied in the literature\cite{kras1} in the
regime of moderate damping $Q\gtrapprox 5$, and we consider here in
further detail application of the model in that regime.

It is instructive to define a normalised current-ramp rate
$\Gamma_\textrm{I}\equiv{}\frac{1}{I}{\textrm{d}I}/{\textrm{d}t}$.
Eqn.~\ref{pI} can then be rewritten
\begin{equation} \label{GEGI}
p(I)=\frac{\Gamma_\textrm{E}}{\Gamma_\textrm{I}}.\frac{1-\int_0^Ip(I')\textrm{d}I'}{I}.
\end{equation}
For small currents, $\Gamma_\textrm{E} \ll \Gamma_\textrm{I}$, so
$p(I)$ is small. As the current is increased towards the current
$I_\textrm{EI}$, at which $\Gamma_\textrm{E}=\Gamma_\textrm{I}$, the
first quotient in Eqn.~\ref{GEGI} increases and therefore, as the
current increases further, the numerator of the second
quotient\footnote{Note that the numerator is the total probability
of there having been no switch as the current ramps from 0 to $I$,
so that the quotient represents the average probability per unit
current of there having been no switch.} begins to reduce from 1 to
zero. The maximum in $p(I)$ therefore occurs for $\Gamma_\textrm{E}
\gtrsim \Gamma_\textrm{I}$. The dashed lines in
Fig.~\ref{fig:IswthT}b show the currents at which
$\Gamma_\textrm{E}=\Gamma_\textrm{I}$ and
$\Gamma_\textrm{E}=10\Gamma_\textrm{I}$. The exact ratio
$\Gamma_\textrm{E}/\Gamma_\textrm{I}$ at the maximum in $p(I)$ is
temperature dependent: at 5 K, the peak in the switching current
lies at a higher current than the current at which
$\Gamma_\textrm{E}=10\Gamma_\textrm{I}$, whereas at 15 K, the peak
in the switching current lies at a lower current than the current at
which $\Gamma_\textrm{E}=10\Gamma_\textrm{I}$.
\begin{figure}[!hbp]
\includegraphics[width=8.6cm]{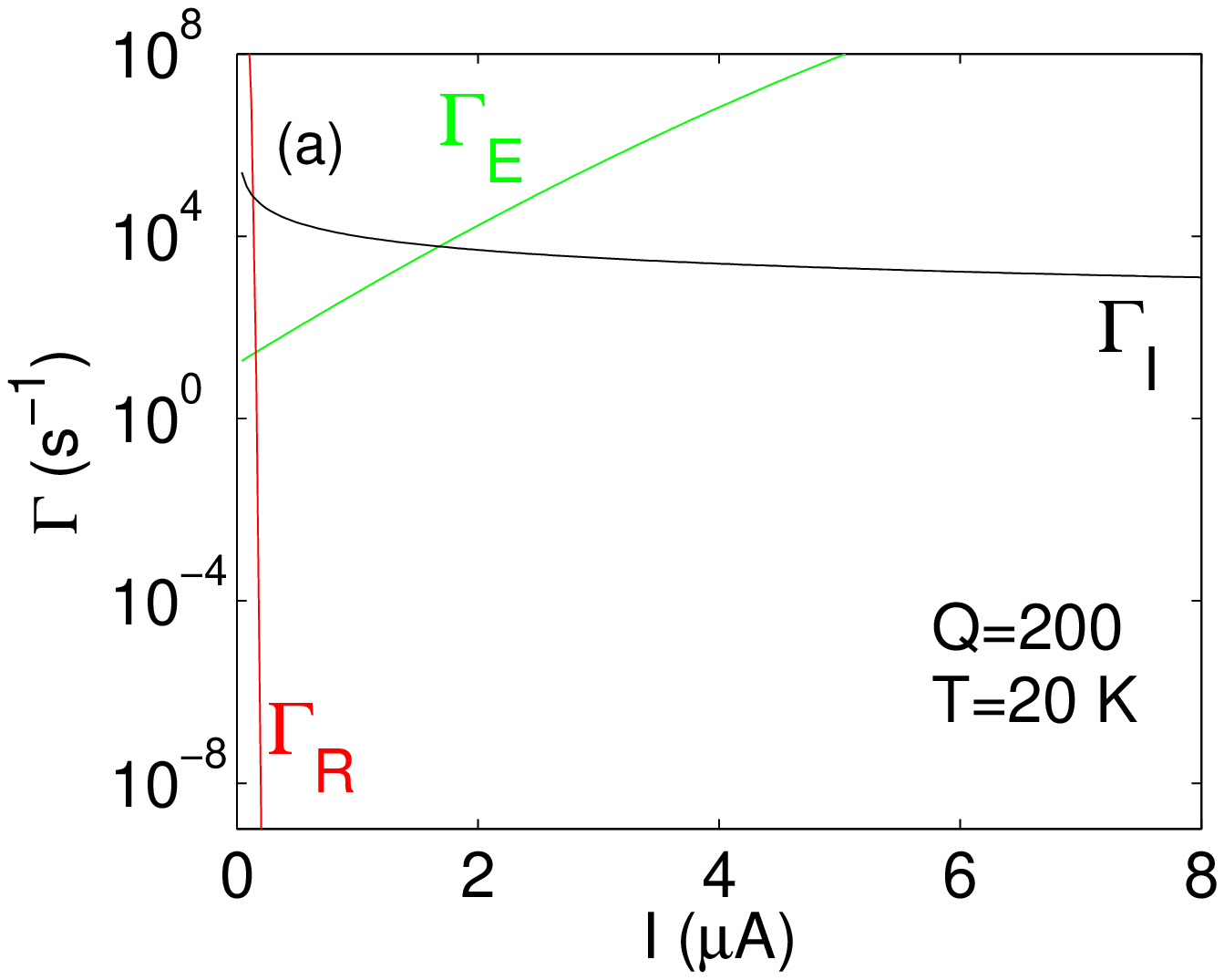}
\includegraphics[width=8.6cm]{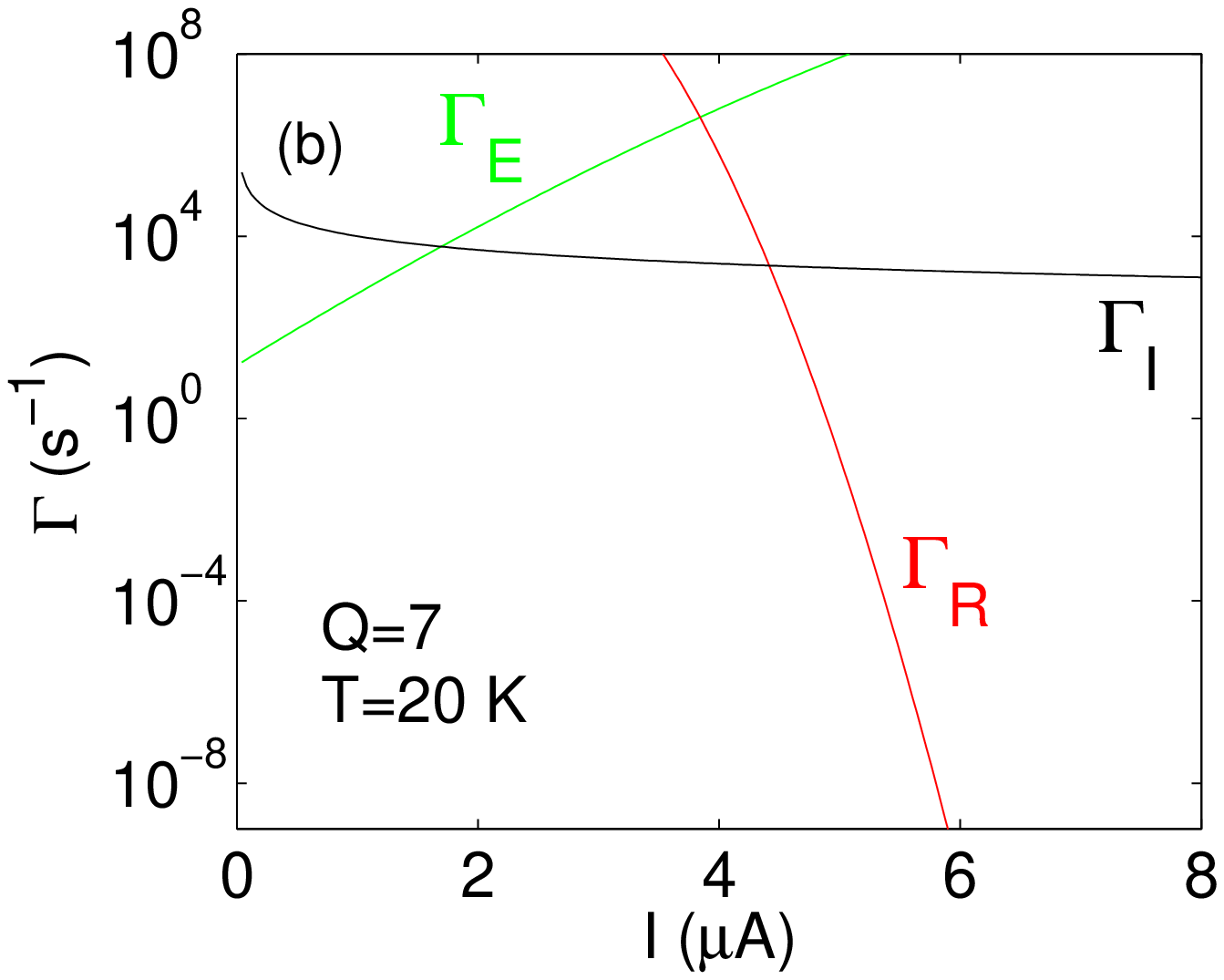}
\includegraphics[width=8.6cm]{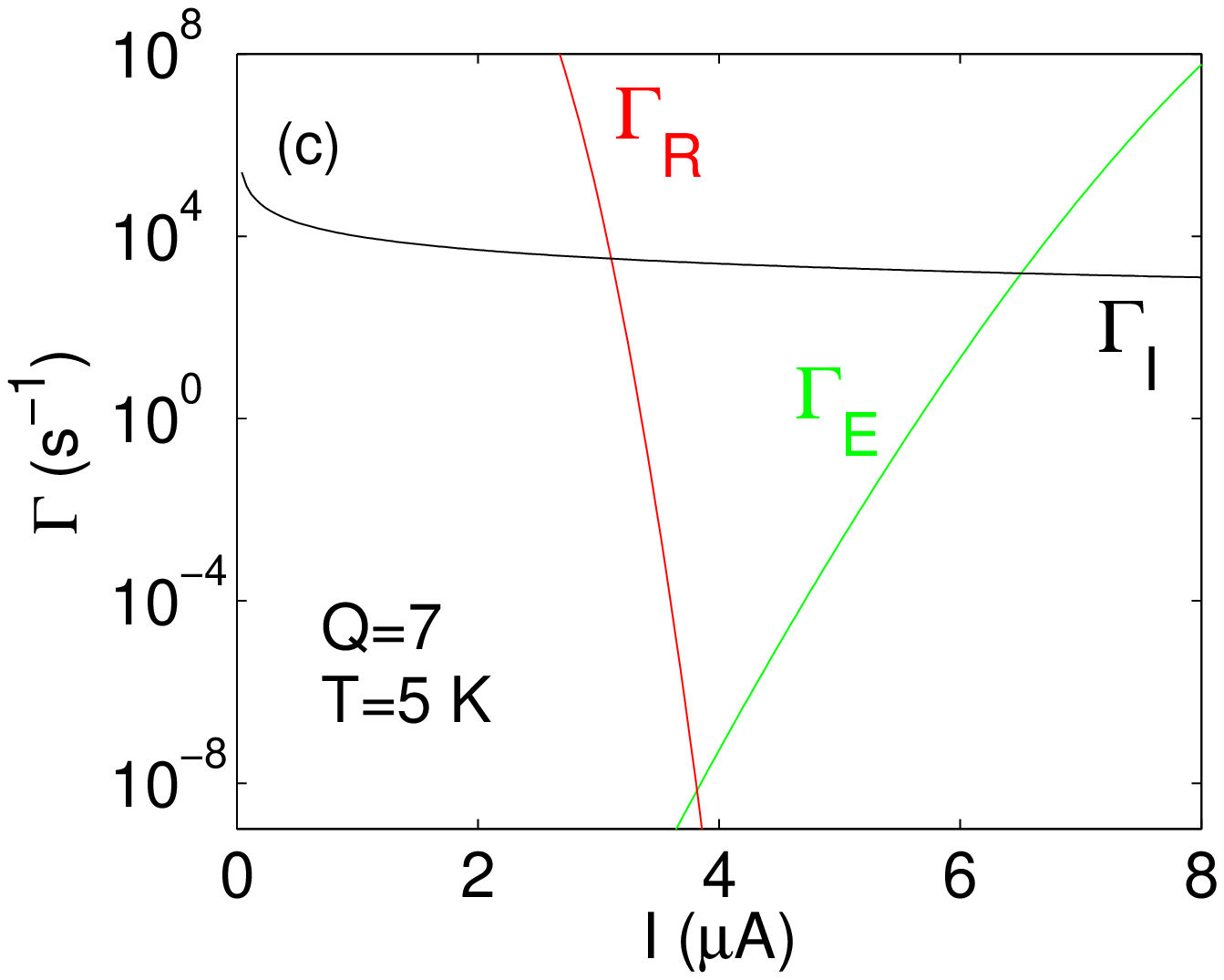}
\caption{(Color online) \label{fig:GhighQ} \label{fig:GmodQlowT}
\label{fig:GmodQ}Variation of characteristic rates with current. The
different panels show the effect of variations in temperature and
$Q$.}
\end{figure}

\section{The multiple switch-retrapping regime}
As the current is increased from zero, the three characteristic
rates $\Gamma_\textrm{E}$, $\Gamma_\textrm{R}$ and
$\Gamma_\textrm{I}$ vary. Fig.~\ref{fig:GhighQ}a shows the variation
of these three rates when $Q=200$, \textit{i.e.}, for an underdamped
junction. At very low currents, the retrapping rate is much larger
than $\Gamma_\textrm{E}$ and $\Gamma_\textrm{I}$. Also, since
$\Gamma_\textrm{E}\ll\Gamma_\textrm{I}$, no escape events occur.
When the current is increased to around $I_\textrm{EI}$, an escape
event becomes likely, but for $I\gtrapprox I_\textrm{EI}$ the
retrapping rate is very much smaller than the escape rate. Therefore
retrapping is negligible in the case illustrated in
Fig.~\ref{fig:GhighQ}a.
As we will see, an important current is the current $I_\textrm{ER}$
at which $\Gamma_\textrm{E}=\Gamma_\textrm{R}$. In
Fig.~\ref{fig:GhighQ}a, $I_\textrm{ER}=0.31$ $\mu$A and
$I_\textrm{ER}<I_\textrm{EI}$.

\begin{figure}[!hbp]
\includegraphics[width=8.6cm]{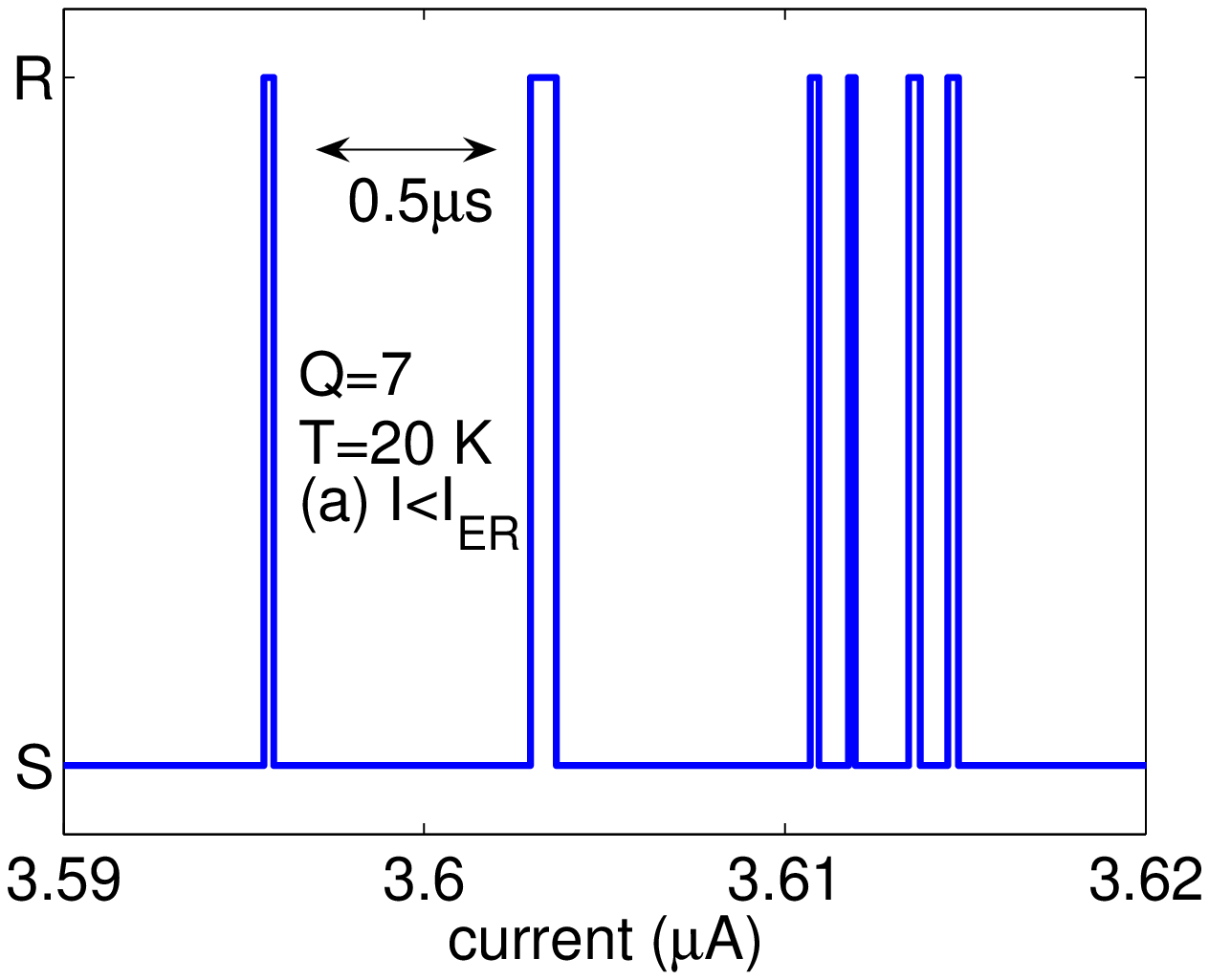}
\includegraphics[width=8.6cm]{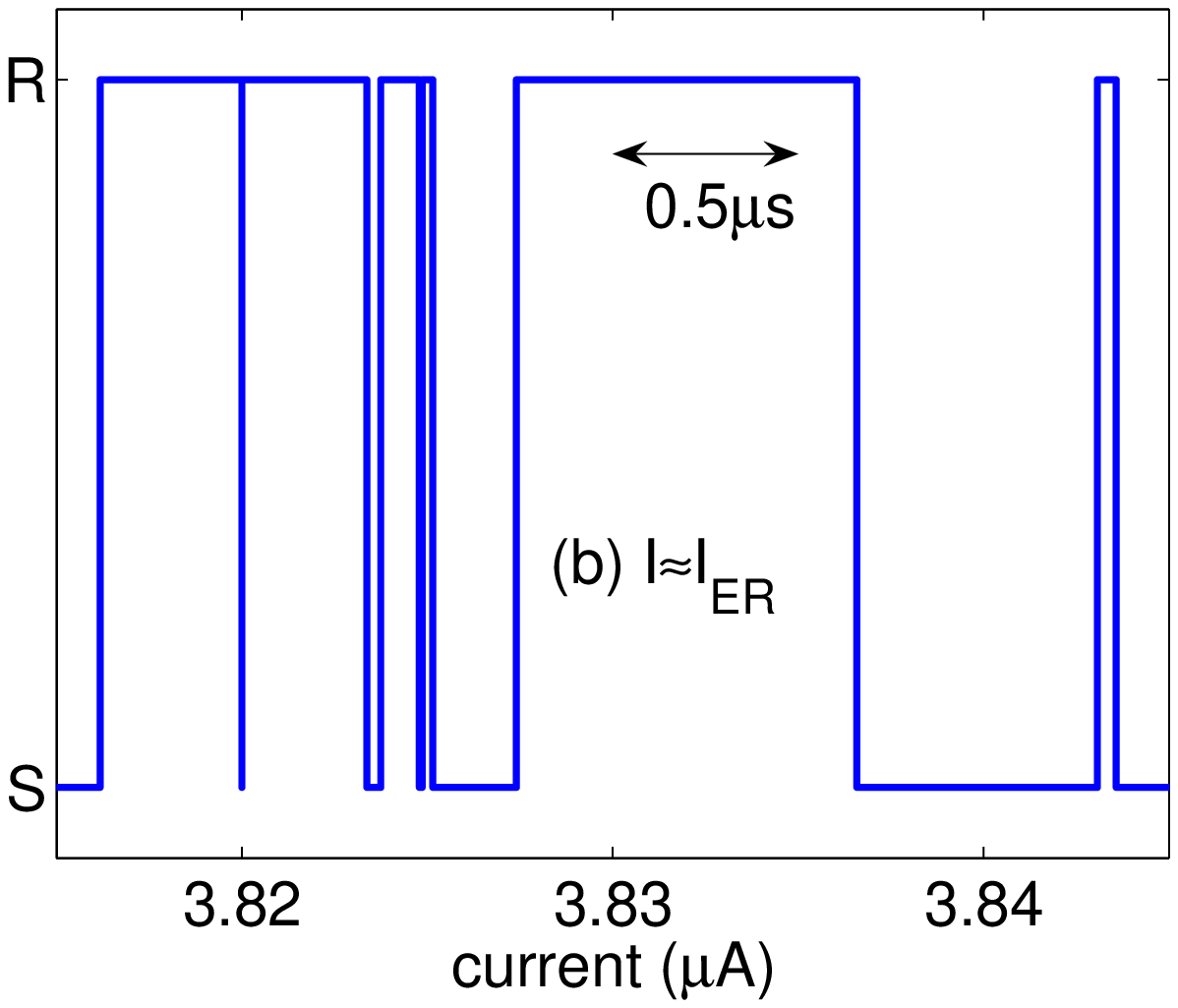}
\includegraphics[width=8.6cm]{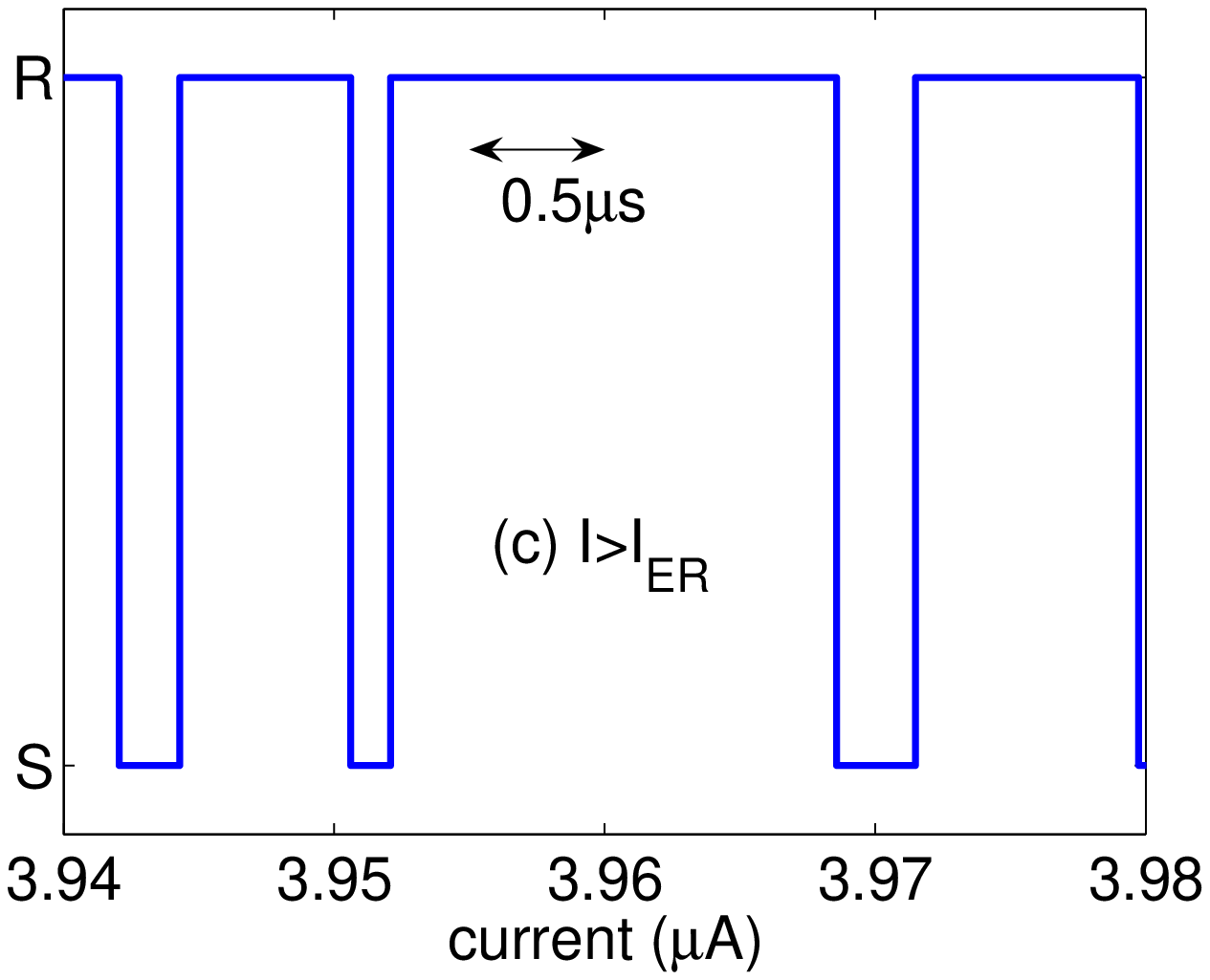}
\caption{(Color online) \label{fig:Vt}Switching between zero-voltage
supercurrent state (S) and the resistive running state (R) at three
currents close to $I_\textrm{ER}=3.8417$ $\mu$A in a representative
simulation for the junction parameters shown in
Fig.~\ref{fig:GmodQ}b.  The instantaneous voltage in the running
state well above the retrapping current is given by $IR$, where $R$
is the relevant resistance shunting the junction.}
\end{figure}

Fig.~\ref{fig:GmodQ}b shows the variation of the three
characteristic rates for a more heavily damped junction. The escape
rate is only weakly dependent on $Q$ through the pre-factor
$a_\textrm{t}$ (Eqn.~\ref{GE}). The retrapping rate is exponentially
dependent on $Q$ (Eqn.~\ref{GR}); it is much larger in
Fig.~\ref{fig:GmodQ}b than in Fig.~\ref{fig:GhighQ}a and
$I_\textrm{ER}>I_\textrm{EI}$. For currents $I\sim I_\textrm{EI}$,
the retrapping rate $\Gamma_\textrm{R}$ is now much larger than
$\Gamma_\textrm{E}$. Escape events occur for $I\gtrapprox
I_\textrm{EI}$, but retrapping occurs shortly afterwards; the
particle moves down the washboard in fits and starts and the
time-averaged voltage across the junction is non-zero --- this state
can be called a region of phase diffusion.\footnote{See also the
discussion in Section \ref{pd}.} As the current increases, the
escape and retrapping rates become more and more similar, so there
is a gradual increase in the time-averaged voltage.
Fig.~\ref{fig:Vt} shows, for the same values of $Q$ and $T$ as
Fig.~\ref{fig:GmodQ}b, a simulation of jumps between the
supercurrent (zero voltage) and running (resistive) states at three
currents close to $I_\textrm{ER}$. In Fig.~\ref{fig:Vt}a,
$I<I_\textrm{ER}$ and the junction spends most of the time in the
zero-voltage state. At $I\sim I_\textrm{ER}$ (Fig.~\ref{fig:Vt}b),
escape events and retrapping events are expected in similar
proportion and the junction spends a similar amount of time in the
zero-voltage and escaped states. The time-averaged voltage across
the junction becomes a significant fraction of the fully switched
voltage, so an experiment is likely to measure a switch event. As
the current is increased further above $I_\textrm{ER}$, any retrap
event will be followed quickly by an escape event, so the junction
spends almost all its time in the running state, as
Fig.~\ref{fig:Vt}c shows.
For the junction parameters corresponding to Fig.~\ref{fig:GmodQ}b,
the junction switches around $I_\textrm{ER}>I_\textrm{EI}$, so the
switching current is greater than the switching current in the
underdamped case.  In other words, counter-intuitively, thermal
fluctuations suppress the switching current less in the multiple
switching-retrapping regime than in the conventional underdamped
thermally-activated switching regime.\footnote{The counter-intuitive
nature of the accompanying decrease in the width with increasing
temperature has previously been highlighted by Krasnov \textit{et
al.} in Refs.~\onlinecite{kras1} and \onlinecite{kras2}.}

\section{Temperature dependence, the crossover regime and $T^*$}\label{sims}
Figs.~\ref{fig:GmodQlowT}b and c show the variation in the
characteristic rates for two different temperatures with $Q=7$. At
the lower temperature, 5 K, (Fig.~\ref{fig:GmodQlowT}c)
$I_\textrm{EI}>I_\textrm{ER}$ so, for $\Gamma_\textrm{E} \sim
\Gamma_\textrm{I}$, the retrapping rate is smaller than the escape
rate. Therefore, retrapping after escape does not occur, and the
conventional underdamped thermal activation behaviour is obtained.
Conversely, at a higher temperature, 20 K,
(Fig.~\ref{fig:GmodQlowT}b), $I_\textrm{EI}<I_\textrm{ER}$, and so
there are multiple escape and retrapping events, as described
earlier.
Note that $I_\textrm{ER}$ is approximately unchanged as temperature
varies at constant $Q$.  This is expected by inspection of
Eqns.~\ref{GE} and \ref{GR}.  Ignoring corrections of logarithmic
order,
$\Delta{U}_\textrm{E}(I_\textrm{ER})=\Delta{U}_\textrm{R}(I_\textrm{ER})$
 determines $I_\textrm{ER}$, where $\Delta{U}_\textrm{E}$ and $\Delta{U}_\textrm{R}$ are both
 independent of temperature and
 hence $I_\textrm{ER}$ is independent of temperature too.

\begin{figure}[!hbp]
\includegraphics[width=8.6cm]{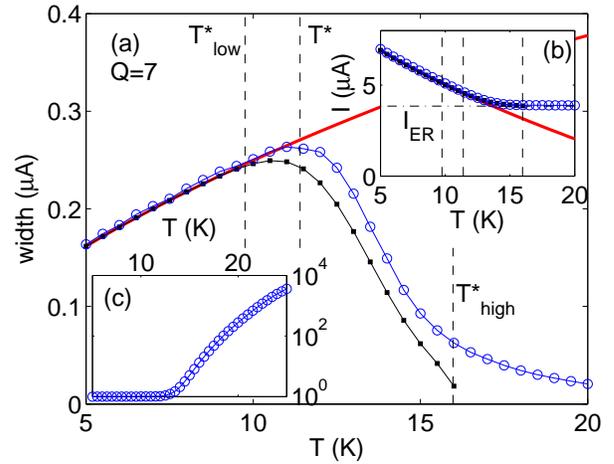}
\caption{(Color online) \label{fig:sigmaT}\label{fig:IswT} (a)
Variation of standard deviation of switching distribution for $Q=7$
with temperature. (b) Variation of mean switching current with
temperature for $Q=7$. Shown are simulation results (open circles,
with line to guide the eye, blue online), the full underdamped
thermal (thick line, red online) and the underdamped thermal
distribution truncated to above $I_\textrm{ER}$ (small closed
circles, with black line to guide the eye). The dash-dotted line
shows $I_\textrm{ER}$ and $T^*$ is defined by the maximum in the
simulated distribution width. (c) \label{fig:Nsw}Variation with
temperature of the mean number of escape events before a switch is
counted in simulations. In the underdamped thermal case, a single
escape event would lead to a switch. Above $T^*_\textrm{high}$, the
distribution in this value is smaller than the symbols.  Simulated
distributions were based on 25000 switching events.}
\end{figure}
\begin{figure}[!hbp]
\includegraphics[width=8.6cm]{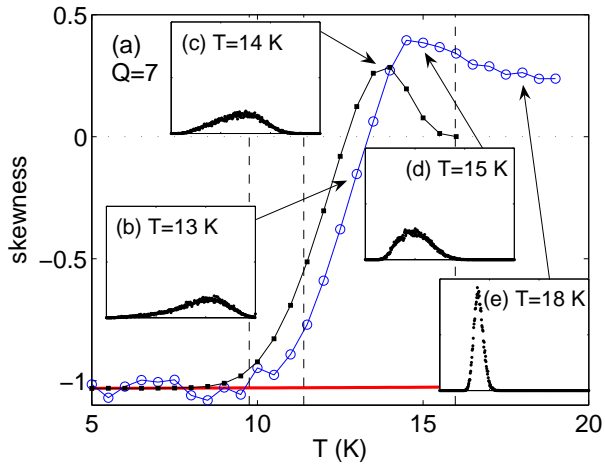}
\caption{(Color online) \label{fig:skewT} (a) The skewness of the
switching distribution, negative for conventional underdamped
distributions, becomes positive around $T^*_\textrm{high}$. Points
and lines as described for Fig.~\ref{fig:sigmaT}.  (b)--(e)
Switching distributions at selected temperatures. Note the scales on
these insets are the same, with the scale on the horizontal current
axis running between 3.5 $\mu$A and 5 $\mu$A and the scale on the
vertical $p(I)$ axis running from 0 to 10 $(\mu$A$)^{-1}$.
 Simulated distributions were based on 25000 switching events.}
\end{figure}
The results of simulations of these dynamics over a broader range of
temperatures are shown in Fig.~\ref{fig:sigmaT}.  In
Fig.~\ref{fig:sigmaT}a, the width of the distribution follows the
conventional underdamped thermal behaviour ($\sigma \sim T^{2/3}$)
at lower temperatures, passes through a maximum and then falls at
higher temperatures, matching experimental observations. In the
previous experimental reports,\cite{kivioja,mannik,kras1} a
characteristic temperature $T^*$ was defined as the temperature at
which the maximum in $\sigma(T)$ occurs.
At low temperatures, a single escape event results in a switch being
counted (Fig.~\ref{fig:Nsw}c) and the mean of the distribution
follows the conventional underdamped thermal behaviour
(Fig.~\ref{fig:IswT}b). At around $T^*$, the mean switching current
flattens out and reaches an approximately constant value
$I\approx{}I_\textrm{ER}$ well above $T^*$.  For higher temperatures
a significant number ($\sim{10^3-10^4}$ above 25 K) of escape events
occurs before a switch is counted.
 The shape of the switching distribution also
changes as the temperature is increased.  Fig.~\ref{fig:skewT}b--e
shows that the shape departs from that shown in Fig.~\ref{fig:pIt}.
The skewness (the ratio of the third moment about the mean to the
standard deviation) gives a simple one-parameter description of the
shape of the distribution; a symmetrical distribution has zero
skewness. The skewness of the underdamped thermal distribution is
around $-1$ over the range of temperature shown in
Fig.~\ref{fig:IswT}. Fig.~\ref{fig:skewT}a shows the variation of
the skewness of the simulated distribution around $T^*$.  The
skewness of the distribution begins to depart from its thermal value
somewhat below $T^*$, becoming progressively less negatively skewed
and then positively skewed, passing through a maximum and then
beginning to level out at around the same temperature as the width
begins to level out and as the mean switching current levels off.

From these simulations, we identify three different regimes of
behaviour.  At low temperatures, conventional thermal underdamped
behaviour is observed.  At some higher temperature below $T^*$, the
skewness of the distribution, and in detail also the width and the
mean, depart from the underdamped thermal values.  Above this
temperature, the skewness and width vary rapidly.  At a higher
temperature, there is a crossover to a different regime in which the
mean switching current is approximately constant and the skewness
and width are slowly decreasing as the temperature is increased.  To
describe this behaviour, we label the two boundaries between these
three regimes $T^*_\textrm{low}$ and $T^*_\textrm{high}$, where
$T^*_\textrm{low}<T^*<T^*_\textrm{high}$.

\begin{figure}[!hbp]
\includegraphics[width=8.6cm]{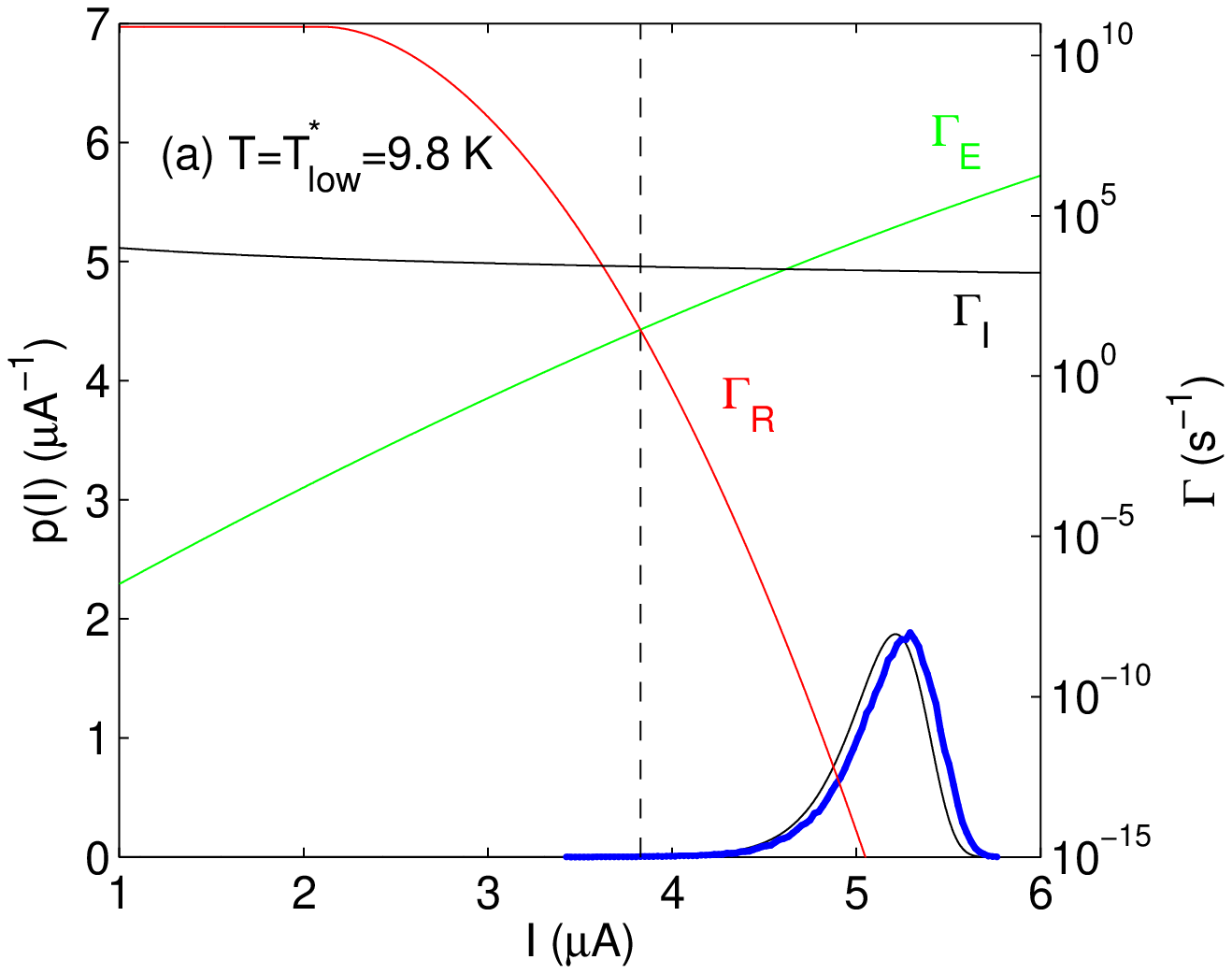}\\
\includegraphics[width=8.6cm]{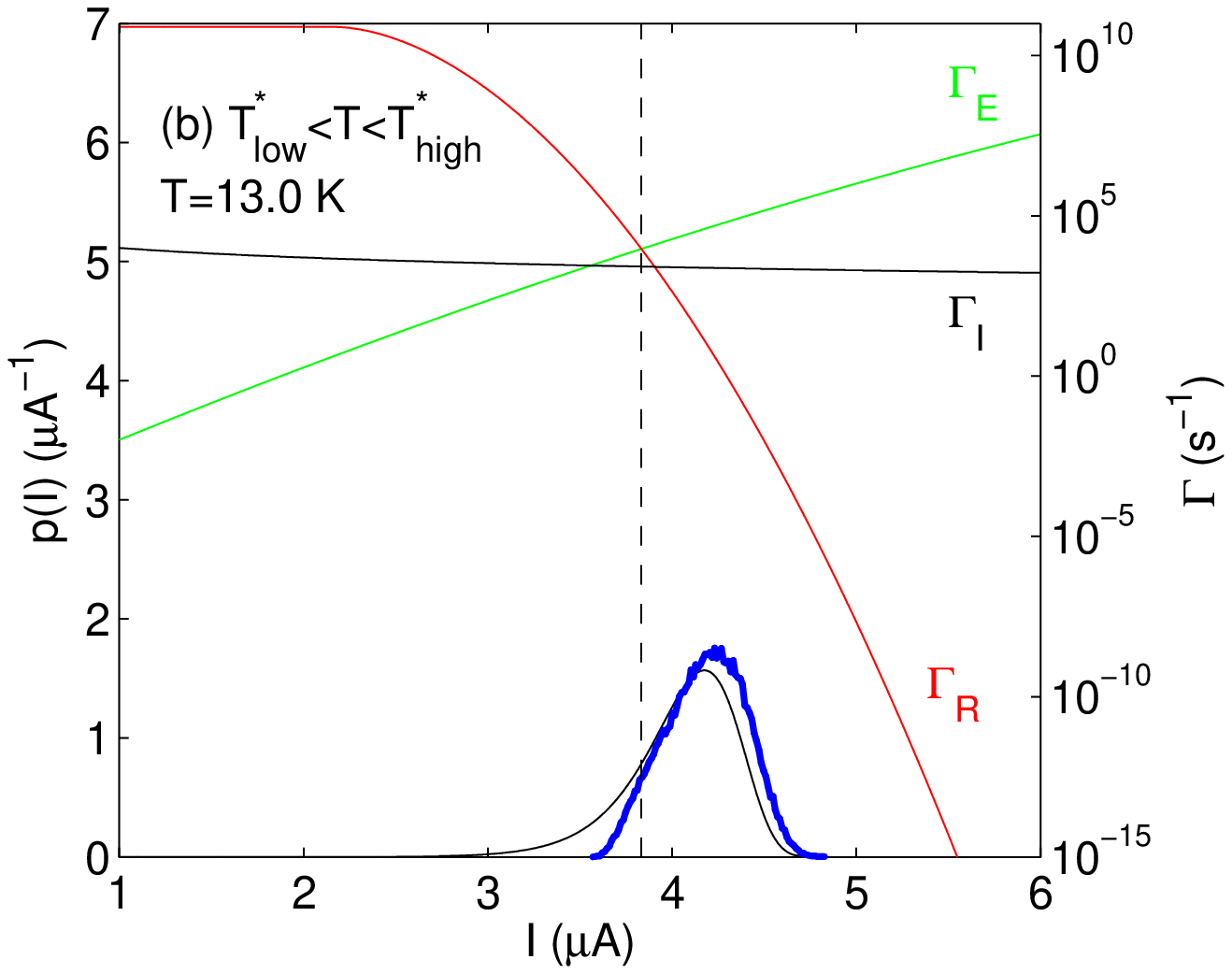}\\
\includegraphics[width=8.6cm]{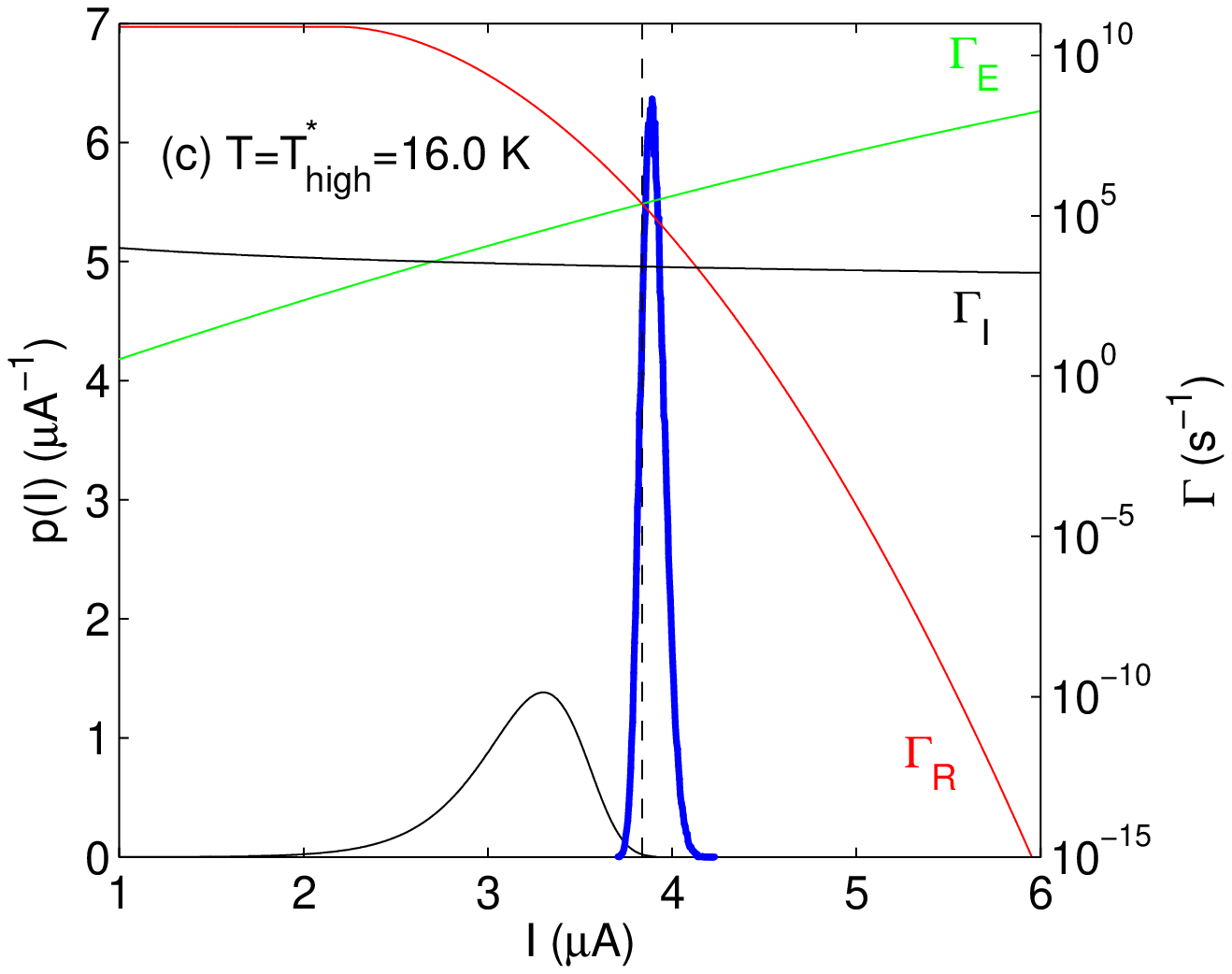}
\caption{(Color online) \label{fig:Tstars} Variation of
characteristic rates with current at three temperatures for $Q=7$,
along with the switching distribution. The heavy (blue online)
points show the simulated switching distribution. The lower black
curve shows the underdamped thermally activated switching
distribution. The broken black line depicts the current
$I_\textrm{ER}$.  (a) $T=T^*_\textrm{low}=9.8$ K; $I_\textrm{ER}$
coincides with the bottom of the thermal distribution.  (b) $T=13.0$
K. (c) $T=T^*_\textrm{high}=16.0$ K; $I_\textrm{ER}$ coincides with
the top of the thermal distribution. Simulated distributions were
based on 100000 switching events.}
\end{figure}

To arrive at a quantitative definition for $T^*_\textrm{low}$, we
note that retrapping only has a significant effect on the dynamics
when there are escapes at currents $I<I_\textrm{ER}$. Therefore, for
parameters where there are no escapes for $I<I_\textrm{ER}$, the
switching distribution does not depart from the conventional
underdamped thermal distribution; we define $T^*_\textrm{low}$
quantitatively as the temperature at which $I_\textrm{ER}$ coincides
with the bottom of the conventional thermally activated underdamped
switching distribution (see also the lower line (pink online) in
Fig.~\ref{fig:IswthT}), where we define the bottom $I_\textrm{b}$
and top $I_\textrm{t}$ of the distribution by
$\int_{0}^{I_\textrm{b}} p(I)\textrm{d}I=f_\textrm{p}$ and
$\int_{I_\textrm{t}}^{I_\textrm{c}} p(I)\textrm{d}I=f_\textrm{p}$,
where $0<f_\textrm{p}\ll 1$, with $f_\textrm{p}=0.0005$.
Fig.~\ref{fig:Tstars}a shows a simulated switching-current
distribution at $T^*_\textrm{low}$ and a comparison of
$\Gamma_\textrm{E}$, $\Gamma_\textrm{R}$ and $\Gamma_\textrm{I}$ as
a function of current.

As the temperature increases, the current $I_\textrm{EI}$ decreases.
For temperatures $T>T_{\textrm{low}}^*$, escape events occur for
$I_\textrm{EI}\lesssim I\lesssim I_\textrm{ER}$ and are followed by
retrapping events; they do not result in the count of a switch.
Fig.~\ref{fig:Tstars}b shows, for a temperature
$T>T_{\textrm{low}}^*$, a comparison of the characteristic rates and
also shows that the simulated switching distribution begins to
depart from the conventional underdamped thermally activated
switching distribution.

For $T>T_{\textrm{low}}^*$, escape events leading to
\textit{switching} only occur at currents $I\gtrsim I_\textrm{ER}$.
When a part of the underdamped thermal distribution lies at
$I\gtrsim I_\textrm{ER}$, one might na\"{\i}vely expect that the
width, mean and shape of the switching distribution would be
approximately the same as the width, mean and shape of the part of
the underdamped thermal distribution lying at $I>I_\textrm{ER}$. The
mean, width and skewness of the part of the underdamped thermal
distribution lying above $I_\textrm{ER}$ (``the truncated
distribution'') are shown in Fig.~\ref{fig:sigmaT} and
Fig.~\ref{fig:skewT}. The mean in the simulations closely matches
the mean of the truncated underdamped thermal distribution, and the
temperature dependence of the width and the skewness in the
simulations also follow the respective variations in the width of
the truncated distribution, up to temperatures approaching the
temperature at which $I_\textrm{ER}$ coincides with the top of the
conventional thermally activated underdamped switching distribution
(see also the upper line (blue online) in Fig.~\ref{fig:IswthT}). We
define this latter temperature as $T_{\textrm{high}}^*$ (see
Fig.~\ref{fig:Tstars}c). The shape of the part of the underdamped
thermal distribution with $I>I_\textrm{ER}$ largely determines the
shape of the switching distribution for
$T_{\textrm{low}}^*<T<T_{\textrm{high}}^*$,\textit{i.e.}, the
conventional thermal behaviour is only followed below
$T^*_\textrm{low}$, but the variations in the mean and width of the
distributions remain analytically determinable for temperatures up
to $T_{\textrm{high}}^*$.

The departure of $I_\textrm{sw}$ from the conventional thermal
activated behaviour above $T_{\textrm{low}}^*$ and the approach to
an asymptotic value is in agreement with experimental results
reported by Franz \textit{et al.}, in which a plateau in
$I_\textrm{sw}(T)$ is observed above a crossover temperature,
followed by a fall at higher temperatures.  In one sample, however,
they observe an increase in $I_\textrm{sw}$ above the crossover
temperature. These observations may be explained by considering
temperature variation in $I_\textrm{c}$ and $Q$. The
Ambegaokar--Baratoff relation\cite{ab} suggests a significant
decrease in $I_\textrm{c}$ for $T>T_\textrm{c}/2$.  A reduction in
$I_\textrm{c}$ as $T$ increases would lead to a fall in the
switching current, whereas a decrease in $Q$ could lead to an
increase in the switching current. One might expect a reduction in
$Q$ at higher temperatures as quasiparticle conductivity increases.
This change will be sample dependent; for a typical IJJ sample, the
resistance may fall by a third from low temperature to
$T_\textrm{c}$. However, if the shunt resistance is dominated by the
environmental impedance, $Q$ is expected to be approximately
$T$-independent.

Since, as the temperature increases above $T^*_\textrm{low}$, the
simulated distribution becomes progressively less negatively skewed
and becomes positively skewed around $T^*_\textrm{high}$, the
temperature variation of the skewness provides a straightforward
experimental way to determine $T^*_\textrm{low}$ and
$T^*_\textrm{high}$. As far as we are aware, no systematic
experiments investigating variations in the shape of the switching
distribution around $T^*$ have yet been reported.

For $T\gtrsim{}T^*_\textrm{high}$, the behaviour is not associated
with the underdamped thermal distribution; we will not discuss this
behaviour in detail.  Fig.~\ref{fig:sigmaT}b shows that, for
$T>T_\textrm{high}^*$, the mean switching current approaches
$I_\textrm{ER}$. Since $I_\textrm{ER}$ is a function of
$I_\textrm{c}$ and $Q$, determination of $I_\textrm{ER}$ from the
switching current at $T\gtrsim{}T^*_\textrm{high}$ provides an
additional experimental probe for the determination of
$I_\textrm{c}$ and $Q$, and a consistency check for derivation of
$Q$ from either $T^*$ or from a ratio of the switching and return
currents. In addition, the expectation that the mean of the
switching distribution at temperatures above $T^*_\textrm{high}$
does not vary with temperature for constant $Q$ means that this
measurement might be used as an experimental probe for whether $Q$
is varying with temperature. We return later to discuss how the
presence of frequency-dependent damping affects this plateau.

\label{pd}For $T>T^*_\textrm{low}$, the behaviour shows some
similarities to the extensively studied phenomenon of phase
diffusion, but also some differences.  The time-averaged voltage
across the junction is finite below the switching current, as
expected for conventional phase diffusion.  However, for
$I<I_\textrm{EI}$, the time-averaged voltage across the junction is
identically zero with a high probability. \footnote{Note that, even
above $T^*_\textrm{high}$, zero voltage is expected for
$I<I_\textrm{EI}$ and so the low-bias phase-diffusion voltage
remains zero.} This behaviour contrasts with a phase-diffusion
regime considered by Ivanchenko and Zil'berman\cite{iz1} and others
in which a finite phase-diffusion resistance persists even at
currents $I\rightarrow{0}$. For $I>I_\textrm{EI}$, the time-averaged
junction voltage is finite, but this phase-diffusion regime also
differs physically from that analysed by Ivanchenko and
Zil'berman\cite{iz1}. Ivanchenko and Zil'berman considered that each
escape would lead to a phase shift of only $2\pi$, whereas in our
model the phase shifts are of order
$\omega_\textrm{P}/\Gamma_\textrm{R}\gg{1}$ --- Kautz and Martinis
previously showed that the presence of multiple-$2\pi$ phase
shift--escape events has an appreciable effect on the
phase-diffusion voltage.\cite{kautz} This is associated with the
time $\sim{}1/\omega_\textrm{P}$ needed after the energy barrier is
exceeded for the instantaneous voltage to increase from zero to its
steady value.
In previous experiments on moderately damped junctions, Krasnov
\textit{et al.}\cite{kras1} did not observe any phase-diffusion
voltage above $T^*$ until well above $T^*$. This appears to conflict
with our understanding from our analysis.  However, the explanation
might simply be that the phase-diffusion voltage was too small to be
measurable.

\subsection{Ramp-rate dependence}
\begin{figure}
\includegraphics[width=8.6cm]{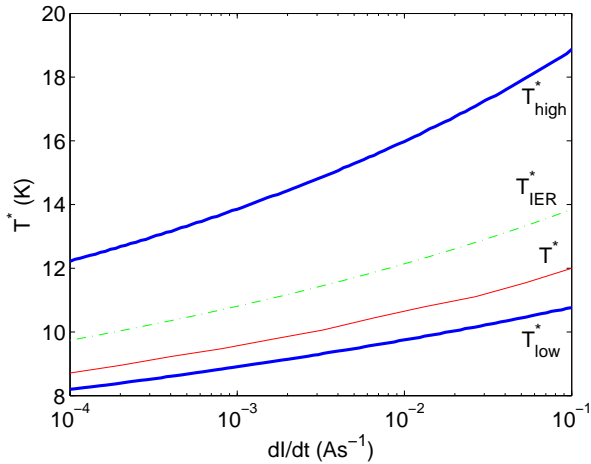}
\caption{(Color online) \label{fig:Tstarcalc}Calculated variation in
$T^*_\textrm{low}$ and $T^*_\textrm{high}$ with ramp rate, with
$f_\textrm{p}$=0.0005 and $T^*_{IER}$, the temperature at which
$\Gamma_\textrm{E}$, $\Gamma_\textrm{R}$ and $\Gamma_\textrm{I}$ are
equal at some value of current. Also shown is the temperature where
the maximum width of the truncated ($I>I_\textrm{ER}$) distribution
lies.}
\end{figure}
\begin{figure}
\includegraphics[width=8.6cm]{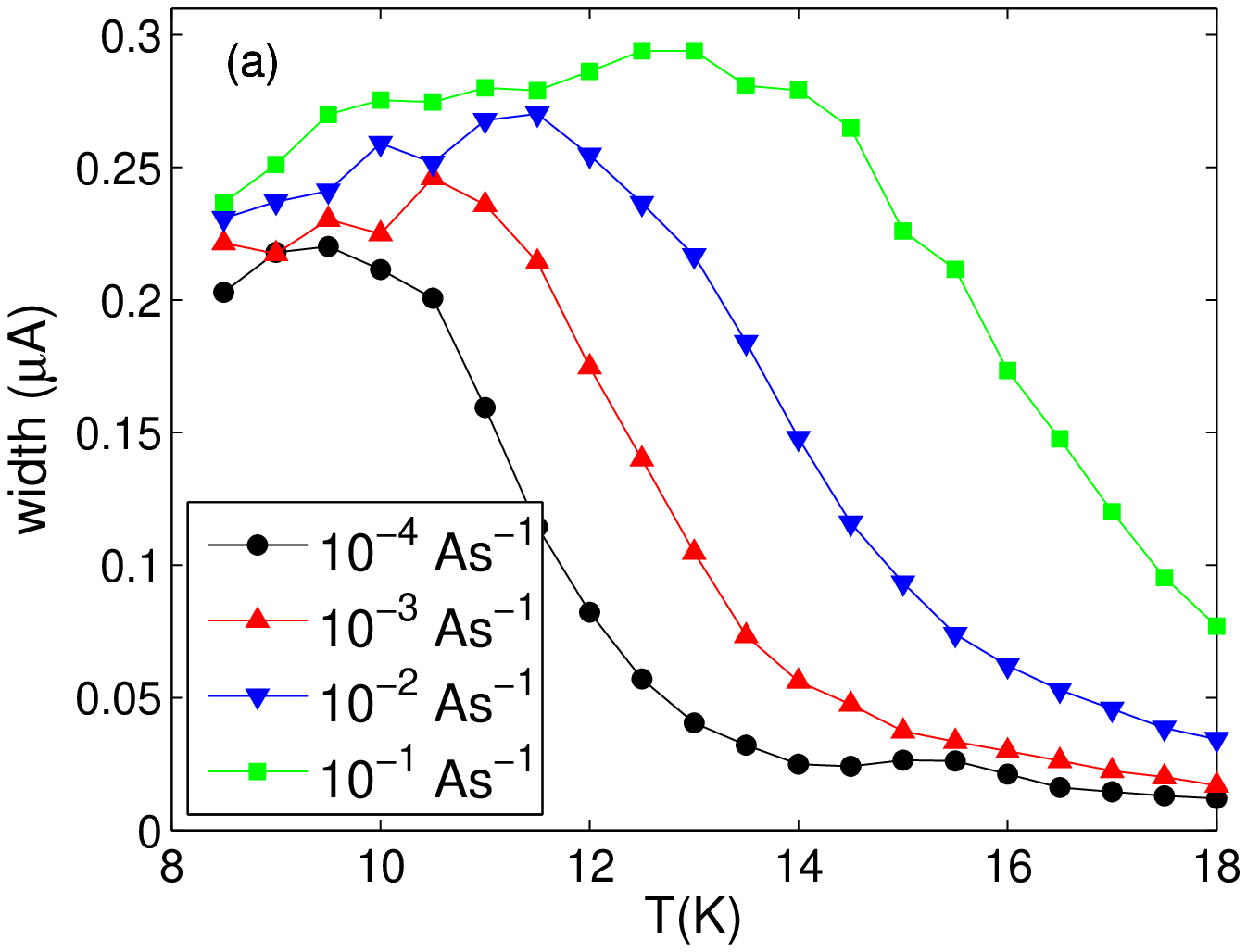}
\includegraphics[width=8.6cm]{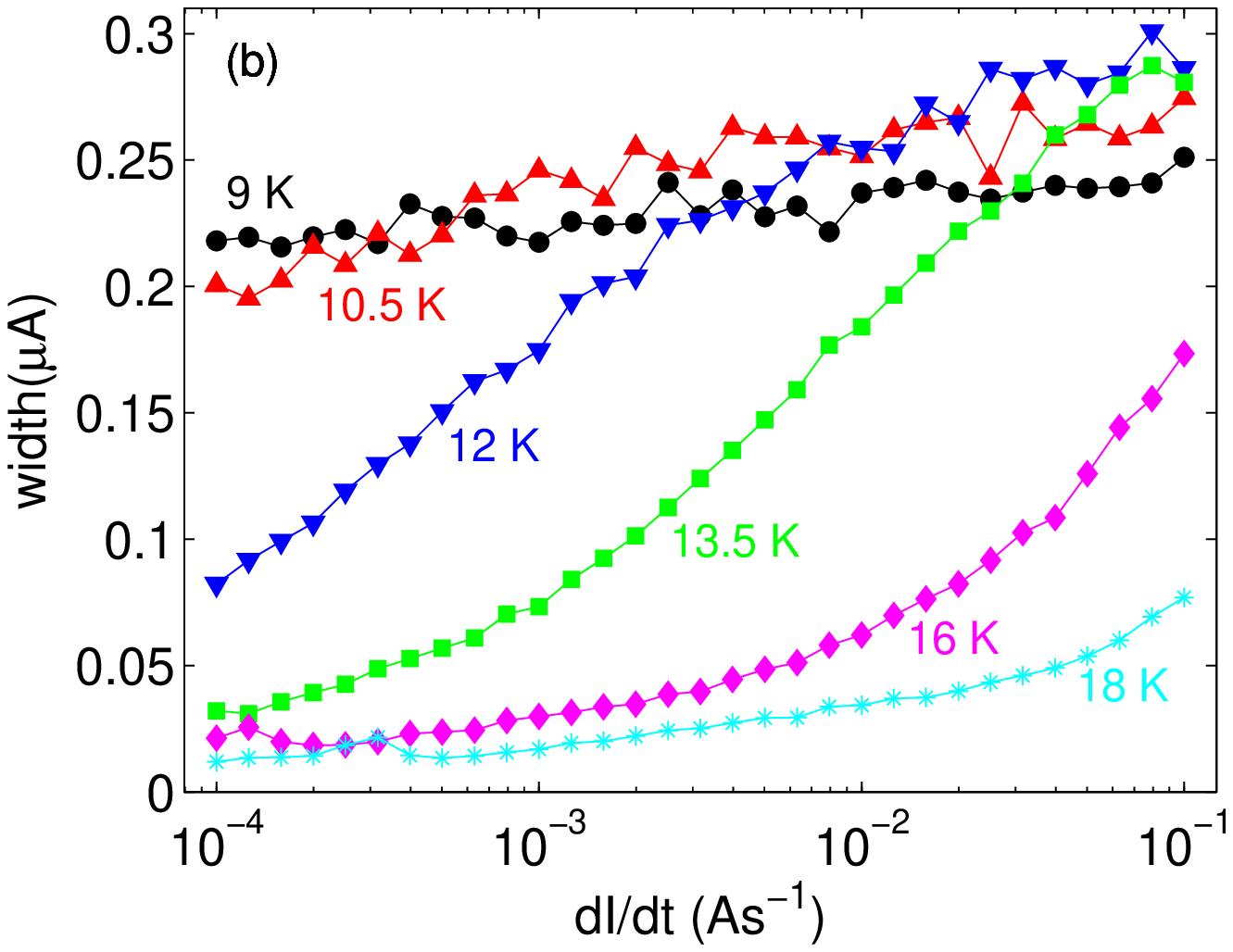}
 \caption{(Color online) (a)\label{fig:TstarsigmaT}
Variation of width with temperature around $T^*$ for various ramp
rates. (b)\label{fig:TstarsigmaIdot} Variation of width with ramp
rate for various temperatures close to $T^*$. Each datapoint
corresponds to a simulated distribution based on 1000 switching
events. Departures from a smooth trend are visible as a result of
statistical fluctuations. Lines are guides to the eye.}
\end{figure}
Since the crossover temperatures are dependent on the shape of the
thermal distribution, and the shape of the thermal distribution is
dependent on the current-ramp rate, the crossover temperatures are
also dependent on the current-ramp rate.  Figure \ref{fig:Tstarcalc}
shows the variation with current-ramp rate of calculated values of
$T^*_\textrm{low}$ and $T^*_\textrm{high}$ and the temperature $T^*$
at which the maximum in the width occurs.  All these values were
determined from the truncated thermal distribution.
Also shown is $T^*_\textrm{IER}$, the temperature at which
$I_\textrm{ER}=I_\textrm{EI}$, in this analysis the most natural
definition for a single crossover temperature.
For a ramp rate $\textrm{d}I/\textrm{d}t=10^{-4}$ As$^{-1}$, we find
$T^*_\textrm{low}=8.2$ K and $T^*_\textrm{high}=12.2$ K, whereas for
a ramp rate $\textrm{d}I/\textrm{d}t=10^{-1}$ As$^{-1}$, we find
$T^*_\textrm{low}=10.8$ K and $T^*_\textrm{high}=18.9$ K.  Thus, as
the simulations reported in Fig.~\ref{fig:TstarsigmaT} show, varying
the ramp-rate can have a significant effect on $T^*$; this
dependence was not recognised in previous reports of the crossover.
Fig.~\ref{fig:TstarsigmaT}a shows simulations of the variation in
the width of the switching distribution with temperature for a
number of current-ramp rates.  In the range shown, for
$\textrm{d}I/\textrm{d}t=10^{-4}$As$^{-1}$ the temperature is above
$T^*_\textrm{low}$, whereas for
$\textrm{d}I/\textrm{d}t=10^{-1}$As$^{-1}$ the temperature ranges
from below $T^*_\textrm{low}$ to above $T^*_\textrm{high}$. This
difference between the two ramp rates leads to a marked difference
in the temperature variation of the width.

Experimentally, probably the most straightforward measurement to
make to investigate these effects would be to keep $T$ fixed and
vary $\textrm{d}I/\textrm{d}t$. Fig.~\ref{fig:TstarsigmaT}b shows a
simulation of this procedure --- there may be a marked difference in
the variation of the width with ramp rate depending on the
temperature of the measurement.  For $T=9$ K, the temperature is at
or below $T^*_\textrm{low}$ for the whole range of ramp rates,
whereas for $T=18$ K, the temperature is at or above
$T^*_\textrm{high}$ for the whole range of ramp rates. For
intermediate temperatures, increasing the ramp rate from $10^{-4}$
As$^{-1}$ to $10^{-1}$ As$^{-1}$ moves the $T^*$ values so that the
temperature is close to $T^*_\textrm{high}$ at the lowest ramp rate
and close to $T^*_\textrm{low}$ at the highest ramp rate.
This ramp-rate dependence of the width could be used to determine
$Q$. As the ramp-rate is varied at a constant temperature, $Q$
remains fixed, but $T^*$ varies and so the width of the distribution
varies, particularly when $T^*$ becomes close to the experimental
temperature. This allows $Q$ to be determined, at each experimental
temperature, by fitting to simulations.

\section{The return current $I_\textrm{r}$, hysteresis and
frequency-dependent damping}
\begin{figure}[!hbp]
\includegraphics[width=8.6cm]{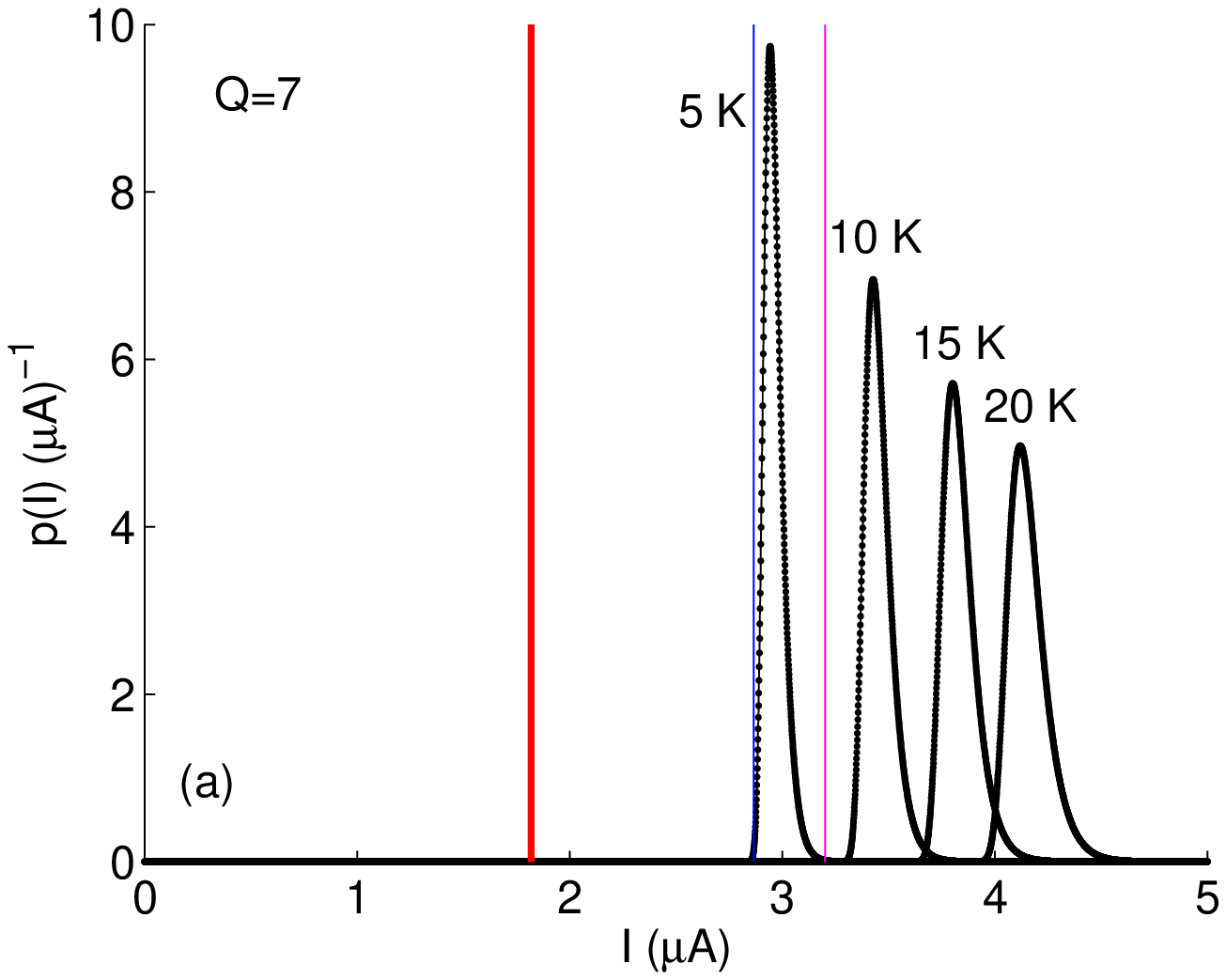}
\includegraphics[width=8.6cm]{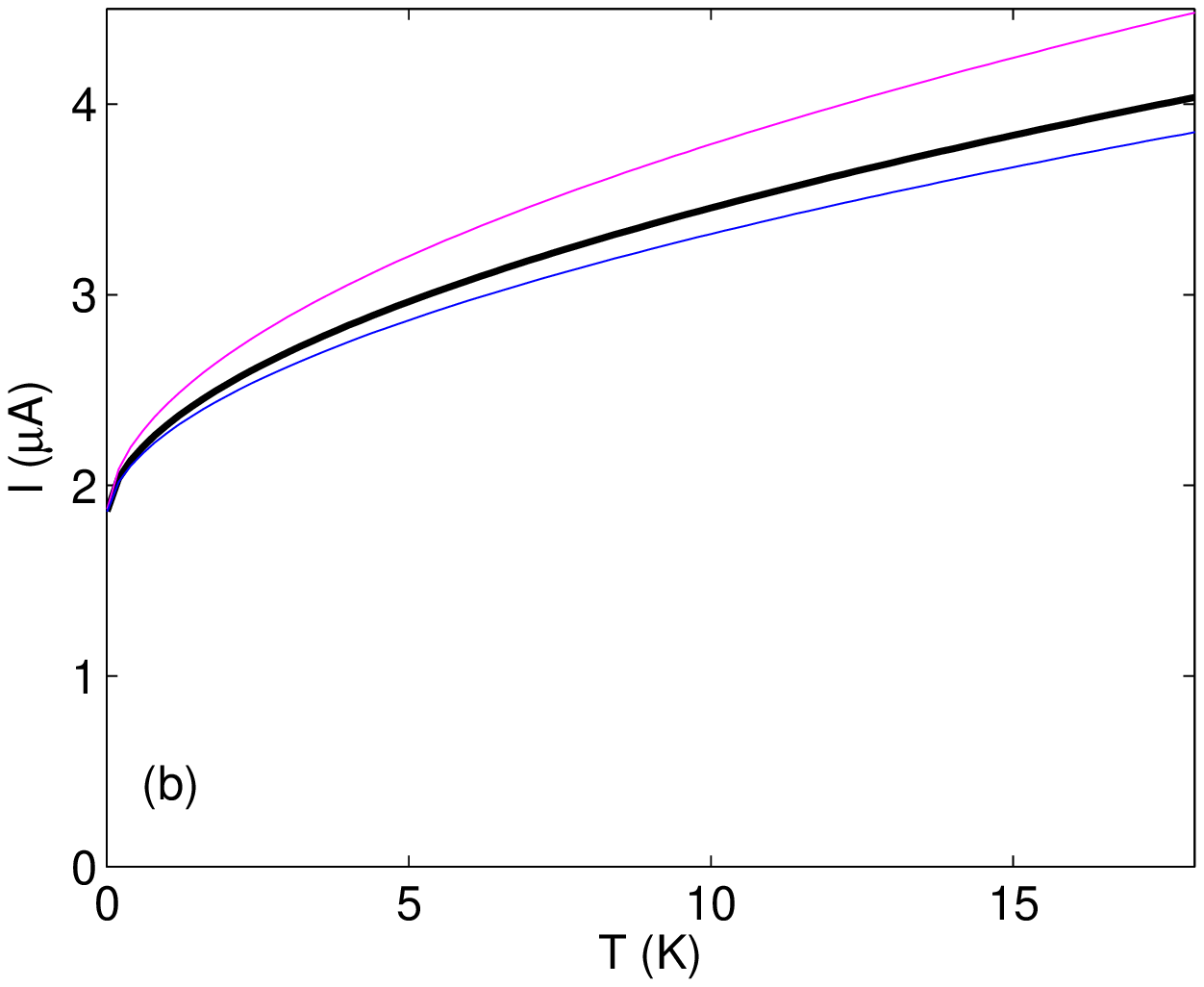}
\caption{(Color online) \label{fig:pRt} (a) Thermally activated
return distributions at various temperatures, for $Q=7$. The
distribution was constructed from the retrapping rate in
Eqn.~\ref{GR} in the same way that Eqn.~\ref{GE} was used to
construct the $p(I)$ distribution shown in Fig.~\ref{fig:pIt}.
Vertical lines (pink and blue online) show the boundaries of the 5 K
distribution, within which 99.99\% of switching events occur. The
thick vertical line (red online) shows the return current in the
absence of fluctuations. \label{fig:IRthT} (b) Variation with
temperature of the mean return current (thick black line) and the
top (pink online) and bottom (blue online) of the return current
distribution.}
\end{figure}

In previous publications\cite{kras1,franz} it was noted that,
although the temperature at which the hysteresis in the $IV$
characteristic disappeared was \textit{around} $T^*$, there was some
difference between the two values. Motivated by this discrepancy, we
consider here in more detail the variation of hysteresis around
$T^*$.

In the RCSJ model in the absence of fluctuations, as an applied
current is ramped down from $I_\textrm{c}$ towards zero, return from
the quasiparticle branch occurs at a current
$I_\textrm{R}\approx{}4I_\textrm{c}/\pi{}Q$ for
$Q\gtrapprox{3}$.\cite{zappe}
Thermal fluctuations lead to retrapping when $\Gamma_\textrm{R}\sim
\Gamma_\textrm{I}$, at currents $I_\textrm{r}>I_\textrm{R}$. As
Fig.~\ref{fig:pRt} shows, there is some distribution in the value at
which return occurs, and the mean and peak of the distribution lie
below the current at which $\Gamma_\textrm{R}=\Gamma_\textrm{I}$.
The width of the return distribution may be shown to vary with $T$
as $\sigma\sim T^{1/2}$.\footnote{Additionally, in a theoretical
article, Chen \textit{et al.}\cite{chen} showed that, close to the
fluctuation-free return current, the voltage departs from $V=IR$,
but we neglect that dependence here.} Experimentally, measurements
of return distributions are more likely than switching-distribution
measurements to be affected by heating and this complicates analysis
of the temperature dependence. To our knowledge, the only report in
the literature of an experiment in which the distribution of return
currents was measured has been given by Castellano \textit{et
al.}\cite{castellano}

\begin{figure}[!htp]
\includegraphics[width=8.6cm]{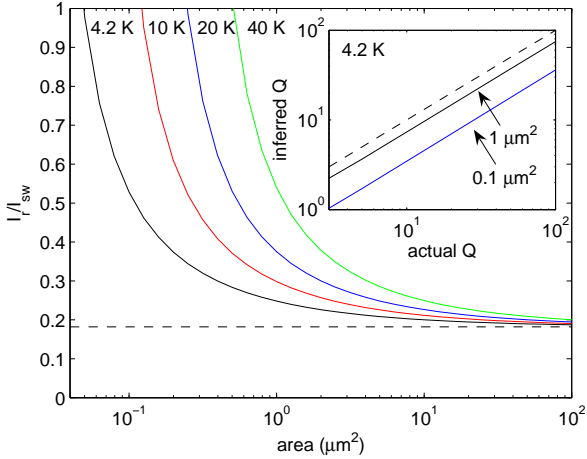}
\caption{(Color online) \label{fig:IrIswA} Area dependence of the
ratio of mean return current and mean switching current for a
quasi-dc measurement $\textrm{d}I/\textrm{d}t=10^{-7}$ As$^{-1}$.
The dashed line shows the fluctuation-free value
$I_\textrm{r}/I_\textrm{sw}=I_\textrm{R}/I_\textrm{c}=4/\pi{Q}$.
Inset: Variation of inferred $Q=4I_\textrm{sw}/\pi{}I_\textrm{r}$ at
4.2 K with $Q$ for two junction areas.  The dashed line shows
$Q_\textrm{inferred}=Q_\textrm{actual}$.}
\end{figure}

\begin{figure}[!htp]
\includegraphics[width=8.6cm]{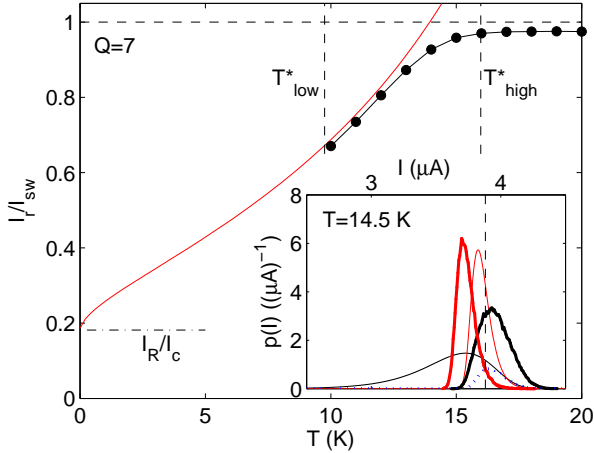}
\caption{(Color online) \label{fig:IrIswT} Temperature dependence of
the ratio of mean return current and mean switching current for
$\textrm{d}I/\textrm{d}t=10^{-2}$ As$^{-1}$. The gray line (red
online) shows the variation derived from the conventional thermal
activation (Figs.~\ref{fig:pIt} and \ref{fig:pRt}) for $Q=7$. Black
points show the dependence obtained from simulations including both
escape and retrapping. The dashed horizontal line shows the value at
which hysteresis disappears $I_\textrm{r}/I_\textrm{sw}=1$.
\label{fig:pIpR} Inset: Simulated switching distribution with
current ramping up (heavy black curve) and simulated return
distribution with current ramping down (heavy gray curve --- red
online), for $T$=14.5 K, demonstrating hysteresis in the switching
and return currents. For comparison, the corresponding conventional
thermally activated distributions are also shown (black and gray
(red online) thin lines). The broken vertical line shows the
position of $I_\textrm{ER}$ and the dotted line (blue online) shows
(the unrescaled) $p_\textrm{S}P_\textrm{nR}(I)$ from
Ref.~\onlinecite{kras1} for comparison. Compare with the
fluctuation-free values, $I_\textrm{c}=10 \mu$A and
$I_\textrm{R}=4I_\textrm{c}/\pi{Q}=1.82$ $\mu$A.}
\end{figure}

A measure of the hysteresis of the junction is given by
$I_\textrm{r}/I_\textrm{sw}$ and, experimentally\cite{kras1},
$I_\textrm{r}/I_\textrm{sw}$ is sometimes used to infer $Q$ through
the approximation
$I_\textrm{r}/I_\textrm{sw}\approx{}I_\textrm{R}/I_\textrm{c}=4/\pi{Q}$.
However, for moderately damped Josephson junctions, the ratio
$E_\textrm{J}/kT$ is often sufficiently small that thermal
fluctuations cause significant departures of $I_\textrm{r}$ from
$I_\textrm{R}$ and $I_\textrm{sw}$ from $I_\textrm{c}$.
Fig.~\ref{fig:IrIswA} shows the variation of
$I_\textrm{r}/I_\textrm{sw}$ with area at selected temperatures,
assuming junctions remain in the underdamped regime. The junction
parameters chosen might be appropriate for intrinsic Josephson
junctions. For large-area junctions,
$I_\textrm{r}/I_\textrm{sw}\approx{}I_\textrm{R}/I_\textrm{c}$.  For
junctions with area $\sim{}1$ $\mu$m$^2$,
$I_\textrm{r}/I_\textrm{sw}$ is significantly larger than
$I_\textrm{R}/I_\textrm{c}$ and so in this case identifying
$I_\textrm{r}/I_\textrm{sw}$ with $I_\textrm{R}/I_\textrm{c}$ to
infer $Q$ will give a significant underestimate of $Q$.  The inset
in Fig.~\ref{fig:IrIswA} shows the difference between the crudely
inferred $Q$ and the true $Q$ at 4.2 K: the crudely inferred $Q$ is
less than the actual $Q$.  If measurements at higher temperatures
were used, the discrepancy would be larger. Therefore it is
important to account for the reduction by thermal fluctuations of
the switching and return currents. We emphasize that, since we are
treating $Q$ as a temperature-independent quantity, this reduction
in the hysteresis is purely a thermal effect.
 For junctions with area 0.1 $\mu$m$^2$, it can be seen by comparing
Figs.~\ref{fig:IswthT} and \ref{fig:IRthT} that the underdamped
return current and underdamped switching current become similar
around 14 K, and at higher temperatures the underdamped return
current exceeds the underdamped switching current.  This would imply
that there is a current range $I_\textrm{sw}<I<I_\textrm{r}$ where
neither the zero-voltage branch nor the resistive branch is stable
--- this is indeed a feature of the behaviour we are describing in the
temperature range $T>T^*_\textrm{low}$ (for example, see
Fig.~\ref{fig:Vt}). Above $T^*_\textrm{low}$, as we have discussed,
it is necessary to consider, in addition, the effects of multiple
escape and retrapping.
Fig.~\ref{fig:IrIswT} shows the variation with temperature in the
ratio $I_\textrm{r}/I_\textrm{sw}$ of these mean values.  This ratio
was obtained from simulations including the effects of both escape
and retrapping as the current is ramped in either direction. At low
temperatures we find
$I_\textrm{r}/I_\textrm{sw}\rightarrow{4/\pi{}Q}$. As the
temperature is increased, even well below $T^*$,
$I_\textrm{r}/I_\textrm{sw}$ departs significantly from
${4/\pi{}Q}$.
Around $T^*$, the distribution departs from its conventional
thermally activated behaviour. Hysteresis is still present for
$T_{\textrm{low}}^*<T<T_{\textrm{high}}^*$, but to a decreasing
extent as $T$ is increased.
The inset of Fig.~\ref{fig:pIpR} shows the switching and return
distributions for a temperature 14.5 K where
$T_{\textrm{low}}^*<T<T_{\textrm{high}}^*$. The thick black curve
shows that the mean and peak of the switching distribution lie
\textit{above} $I_\textrm{ER}$. The thick gray curve (red online)
shows that the mean and peak of the return distribution lie
\textit{below} $I_\textrm{ER}$ (its width is related to the shape of
the conventional return distribution).
 A difference in the mean switching and
return currents, and so some hysteresis, persists even though the
underdamped return current exceeds the underdamped switching
current, which might be thought to imply the absence of hysteresis.
The persistence in hysteresis above $T^*$ in the experiments of
Refs.~\onlinecite{kras1} and~\onlinecite{franz} is likely to be
attributable to the distinction between $T^*$ and
$T^*_\textrm{high}$
--- we expect hysteresis to persist up to $T\approx T^*_\textrm{high}>T^*$.
For $T>T_{\textrm{high}}^*$, both escape and retrapping occur close
to $I_\textrm{ER}$, so that
$I_\textrm{r}/I_\textrm{sw}\rightarrow{1}$ and hysteresis in the IV
characteristic is small.

We would like to note that the probability of a switch being counted
is not the same as the probability of a single escape not followed
by a retrapping event, since a switch may be preceded by many
escape-retrapping events.  This difference was not appreciated in
the quantitative analysis in Ref.~\onlinecite{kras1}, in which the
latter quantity --- although much smaller than the total probability
of a switch --- was evaluated as a function of the current of the
initial escape (compare the dotted curve and the thick black line in
the inset of Fig.~\ref{fig:IrIswT}) and then rescaled\footnote{The
distribution was scaled by dividing by the total probability of a
switch following a single escape event.} and fitted to experimental
data.

\subsection{Frequency-dependent damping} \label{freqdepdt}

In our treatment so far, we have been assuming that the damping $Q$
is frequency independent.  Different retrapping and return behaviour
may arise when the damping is frequency dependent and we now turn to
discuss these differences.
If the damping of the system is frequency-dependent
--- as is likely to be the case unless isolation resistors close to the junction are included or the shunt resistance is much
less than the free-space impedance --- the system will be
characterized by much stronger damping shortly after escape than in
the steady running state.  This means that overdamped behaviour at
escape might lead to multiple escape-retrapping behaviour shortly
after an initial escape but, once the junction has been in the
running state for some time, it becomes much less strongly damped
and might be characterized by the conventional underdamped dynamics
and so unlikely to be retrapped.

In their simulations, Kautz and Martinis\cite{kautz} consider a
crossover to low-frequency damping once the junction has been mostly
in the running state over a timescale $1/\nu_\textrm{c}$, where
$\nu_\textrm{c}$ is a characteristic frequency, a factor of
$10^2-10^5$ smaller than the plasma frequency.  In their simple
model, which captures the qualitative features of the
frequency-dependent--damping behaviour, well above $\nu_\textrm{c}$
the damping is $Q_1$ and well below $\nu_\textrm{c}$ the damping is
$Q_0<Q_1$.

In the simulations presented here so far, we counted a switching
event if the junction was mostly in the running state over a certain
time period. Experimentally, that time period might be determined by
the response time of the measurement electronics.  In the case of
frequency-dependent damping, the time period is set instead by
$1/\nu_\textrm{c}$, since once the junction has been mostly in the
running state for the characteristic time $1/\nu_\textrm{c}$ and
becomes characterized by the much lighter damping $Q_1$,
$\Gamma_\textrm{R}$ falls and retrapping becomes unlikely. This time
is likely to be much shorter than the response time of the
electronics.

To model, through simulations, the effect of a decrease in the
damping at low frequencies, we neglect retrapping once the junction
has been in the running state for a time $1/\nu_\textrm{c}$.
\begin{figure}[!htp]
\includegraphics[width=8.6cm]{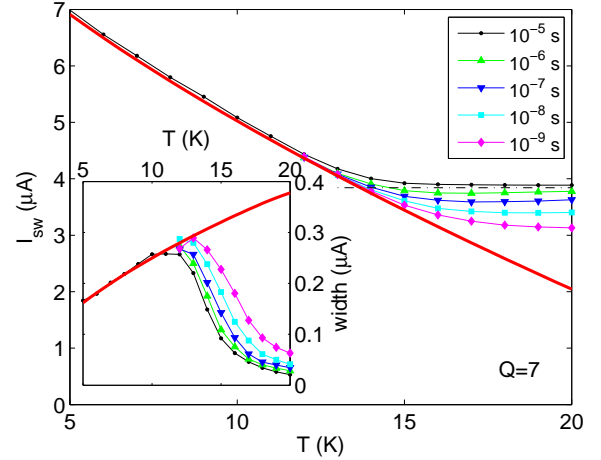}
\caption{(Color online) \label{fig:nuc} Variation of distribution
characteristics with the timescale $1/\nu_\textrm{c}$ for $Q=7$.
Main panel: Mean switching current; the dash-dotted line shows
$I_\textrm{ER}$. Inset: Switching distribution width. Lines joining
points are guides to the eye. The lines without points (red online)
show the conventional underdamped thermal behaviour. Each datapoint
corresponds to a simulated distribution based on 1000 switching
events.}
\end{figure}
Fig.~\ref{fig:nuc} shows the variation in the mean and width of the
switching distribution around $T^*$ as the characteristic time
$1/\nu_\textrm{c}$ varies.
Note that the crossover temperatures $T^*_\textrm{low}$ and
$T^*_\textrm{high}$ are essentially independent of the measurement
time period, since they are set by the crossover between the
extremes of the conventional underdamped distribution and by
$I_\textrm{ER}$, all of which are essentially independent of the
measurement time period.
Fig.~\ref{fig:nuc} shows that a decrease in the characteristic
timescale leads to a decrease in the mean switching current and to
an increase in the width above $T^*$. The width and mean of the
switching distribution for $T>T_{\textrm{high}}^*$ therefore provide
an indication of the crossover frequency $\nu_\textrm{c}$.

This variation with $\nu_\textrm{c}$ may be understood by
considering the relative sizes of $\Gamma_\textrm{E}$ and
$\Gamma_\textrm{R}$, in comparison to $\nu_\textrm{c}$, around the
current at which switching occurs. In the multiple escape-retrap
regime, for small $\nu_\textrm{c}$ such that $\nu_\textrm{c} \ll
\Gamma_\textrm{E},\Gamma_\textrm{R}$, the fraction of time in the
running state $f_\textrm{r}$ is to a very good approximation
$f_\textrm{r,eq}=\Gamma_\textrm{E}/(\Gamma_\textrm{E}+\Gamma_\textrm{R})$
and so switching is likely for
$\Gamma_\textrm{E}=\Gamma_\textrm{R}$, \textit{i.e.}, for
$I=I_\textrm{ER}$. (For the simulations presented in Section
\ref{sims}, the same approximation holds for the temperatures of
interest $T\approx{}T^*$, since the time-period $\tau{}=10^{-5}$ s
$\gg 1/\Gamma_\textrm{E},1/\Gamma_\textrm{R}$ --- compare
Fig.~\ref{fig:Tstars}.) However for larger $\nu_\textrm{c}$ such
that $1/\nu_\textrm{c} \lesssim
(1/\Gamma_\textrm{E}+1/\Gamma_\textrm{R})$, large fluctuations of
$f_\textrm{r}$ away from $f_\textrm{r,eq}$ occur. These fluctuations
make switching likely at $I<I_\textrm{ER}$ and also increase the
width of the switching distribution. The larger $\nu_\textrm{c}$ is,
the smaller $I_\textrm{sw}$ is likely to be. For example, in
Fig.~\ref{fig:Vt}a, for $1/\nu_\textrm{c}=10^{-7}$ s, the escape to
the running state at around 3.603 $\mu$A would last long enough to
cause a switch, although $I_\textrm{ER}=3.8417$ $\mu$A.

The presence of frequency-dependent damping also has a marked effect
on the hysteresis of $IV$ measurements.  For an initial escape as
the current is ramped up from 0, the system is characterized by the
high-frequency damping $Q_0$.  In contrast, when the current is
ramped down from $I_\textrm{c}$, the system is characterized by the
low-frequency damping $Q_1$. The more underdamped behaviour for
return means that hysteresis persists to much higher temperatures
than for a system with frequency-independent damping $Q_0$.

\subsubsection{Application to previous work} \label{discuss}

We now briefly discuss previous work in relation to
frequency-dependent damping.
The work of Krasnov \textit{et al.}\cite{kras1,kras2} is largely on
samples with resistances $\sim{1}$ $\Omega$ which are therefore
likely to be characterized by approximately frequency-independent
damping, although the larger resistance of the IJJ samples suggests
they may be affected by frequency-dependent damping, so that the
damping characterizing retrapping soon after escape is not the same
as the damping extracted from the hysteresis in the $IV$
characteristics.

In their paper, Kivioja \textit{et al.}\cite{kivioja} were
considering mostly the switching dynamics, which are determined by
the high-frequency damping, although the retrapping current
$I_\textrm{m}$ is determined by the low-frequency damping.  They did
not include frequency-dependent damping in their modelling and they
used the high-frequency $R$ as a fitting parameter.
They expect that the maximum in the width of the distribution to
occur at $T_\textrm{d}$, when $\Gamma_\textrm{E}(I_\textrm{m})$
corresponds to their experimental timescale, essentially equivalent
to our $\Gamma_\textrm{I}$. This contrasts with our expectation for
frequency-independent damping that, at $T^*$,
$I_\textrm{sw}=I_\textrm{ER}$.  Since
$\Gamma_\textrm{E}(I_\textrm{m})<\Gamma_\textrm{E}(I_\textrm{ER})$,
the crossover temperature $T^*<T_\textrm{d}$. Kivioja \textit{et
al.} extracted a value for $Q$ from their experimental results which
is therefore larger than would be extracted if thermally activated
retrapping had been included. This may explain the discrepancy
between their extracted $Q=4.4$ and the $Q=4$ suggested by the
nominal values of their experimental parameters.

M\a"annik \textit{et al.}\cite{mannik} obtained values for the
probability of retrapping from numerical Monte Carlo simulations and
included frequency-dependent damping.  This model has the advantage
that it is able to account for the presumably initially increased
rate of retrapping as, in the washboard analog, the particle first
accelerates after escape.  However, the results are less
straightforward to analyse.
The authors expressed the net escape rate as a sum of the
probabilities of multiple escape-retrap events, related to the
thermal escape rate (our Eqn.~\ref{GE}) and the calculated
retrapping probability. Although in detail the assessment of
individual escape probabilities, a nontrivial problem, is
oversimplified and relies on strictly inconsistent
approximations,\footnote{For an escape event involving $n$ retraps
before eventual escape in a time $\Delta{t}$, the average time for
each escape is $\Delta{t}/(n+1)$, implying an equal probability of
each escape event in that average time. However, the probabilities
of the 1st $n$ escapes were set\cite{mannikpc} to 1, with the
$(n+1)$th escape being assigned the probability
$\Gamma\Delta{t}/(n+1)$.  More rigorously, the escape rate could
have been expressed by integrating over all possible values of the
time $t_e$ for each escape, subject to the constraint $\sum
t_e=\Delta{t}$.} the model captures at least qualitatively well the
effect of retrapping on the switching current statistics. M\a"annik
\textit{et al.} found good agreement between their model for the net
escape rate and their experimental results.
Their treatment of the probability of retrapping after escape as a
time-independent quantity contrasts with the model of Ben-Jacob
\textit{et al.}\cite{benjacob}, in which retrapping is modelled by a
rate (Eqn.~\ref{GR}) and so with a probability increasing linearly
with time spent in the running state. The treatment of this
probability as time-independent by M\a"annik \textit{et
al.}\cite{mannik} is successful because, in the model of
frequency-dependent damping which they used, the time over which
retrapping can occur is $\approx 1/\nu_\textrm{c}$. Once the
particle has been in the running state for a time $\gtrsim
1/\nu_\textrm{c}\sim{}1/\omega_\textrm{P}$, low-frequency
(under)damping applies and retrapping is unlikely. This expectation
is borne out by their simulations, in which they observed retrapping
events only times $\lesssim 100/\omega_\textrm{P}$ after
escape.\cite{mannikpc}

\section{Conclusions}
In summary, we have presented discussion and simulations of the
switching and return dynamics of moderately damped Josephson
junctions. We emphasized that there is a regime in which the
junction repeatedly escapes to and retraps from the running state
and demonstrated through the use of simulations that for some
choices of parameters, the number of escapes and retraps during a
single current ramp to an eventual switch into the running state may
be very large ($\sim$10000).  The multiple escape-retrapping regime,
with a large number of escapes of duration
$\sim{}1/\Gamma_\textrm{R}$, is intermediate between the underdamped
regime in which a single escape leads to switching, and the
overdamped phase-diffusion regime in which a very large number of
escapes of very short duration $\sim{}1/\omega_\textrm{P}$ may
occur.

 By examining the region around the crossover in the
temperature dependence of the width in more detail, we showed that
the crossover is, in detail, described by not one but \textit{two}
crossover temperatures $T^*_\textrm{low}$ and $T^*_\textrm{high}$.
The variations in the mean and width of the switching distribution
(in the intermediate regime between the two crossovers) are largely
determined by the shape of the thermally activated switching
distribution and this shape therefore also determines the
temperature of the maximum in the width, the quantity usually
identified as the single crossover temperature.  We showed that the
shape of the switching distribution, parametrized by the skewness,
indicates $T^*_\textrm{low}$ and $T^*_\textrm{high}$. We introduced
a pertinent rate $\Gamma_\textrm{I}$ for understanding the dynamics;
we showed that the details of the frequency dependence of the
junction damping should affect measured values of the mean and width
of switching distribution and weakly affect the crossover
temperatures.

We showed that the characteristic temperatures $T^*$,
$T^*_\textrm{low}$ and $T^*_\textrm{high}$ are all dependent on the
current ramp-rate and therefore are not uniquely determined by
measurements at a particular ramp rate, and in addition that there
is some dependence of the behaviour around and above $T^*$ on any
frequency dependence of the damping.

We also considered the process of return to the supercurrent state
as the current is ramped down in the presence of thermally activated
retrapping events and the implications for measurements of
hysteresis in moderately damped Josephson junctions.  We found that
some hysteresis is expected to persist above $T^*$, to
$T^*_\textrm{high}$, even in junctions with frequency-independent
damping. This suggests a resolution of the issue of hysteresis
somewhat above $T^*$ in previous reports.

\acknowledgments The authors gratefully acknowledge financial
support from the UK EPSRC.

\bibliographystyle{abbrvnat}

\end{document}